\documentclass{pas}
\usepackage{float}
\usepackage[varg]{txfonts}
\usepackage{graphicx}
\usepackage[utf8]{inputenc}

\newcommand{\dcdot}{\mathbin{%
    \nonscript\mspace{-\muexpr\medmuskip*2/3}%
    \cdot
    \nonscript\mspace{-\muexpr\medmuskip*2/3}%
  }%
}

\renewcommand{\figcaptionnumfont}{\bfseries}
\renewcommand{\figcaptiondescfont}{\small}
\renewcommand{\figcaptionfont}{\small} 
\renewcommand{\tablecaptionnumfont}{\bfseries}
\renewcommand{\tablecaptionfont}{\small} 
\renewcommand{\tablefont}{\small}
\renewcommand\footnotesize{\small}

\begin{document}

\lefttitle{Magnetogenesis by galactic processes}
\righttitle{R. Ramesh et al.}

\jnlPage{}{}
\jnlDoiYr{}
\doival{}

\articletitt{Research Paper}

\title{Magnetogenesis by galactic processes: impact on circumgalactic and intergalactic fields}

\author{Rahul Ramesh$^{1,2}$, Enrico Garaldi$^{1,2}$ and Chris Byrohl$^{3}$}

\affil{$^1$Kavli IPMU (WPI), UTIAS, The University of Tokyo, Kashiwa, Chiba 277-8583, Japan \\
$^2$Center for Data-Driven Discovery, Kavli IPMU (WPI), UTIAS, The University of Tokyo, Kashiwa, Chiba 277-8583, Japan \\
$^3$Kavli Institute for Particle Astrophysics \& Cosmology (KIPAC), Stanford University, Stanford, CA 94305, USA}
\corresp{Rahul Ramesh, Email: rramesh@g.ecc.u-tokyo.ac.jp}

\citeauth{}

\history{}

\begin{abstract}
We investigate the origin and evolution of cosmic magnetic fields using a suite of large-volume cosmological magnetohydrodynamic simulations (L$_\mathrm{box}$\,$=$\,$25$ Mpc/h) run with the moving-mesh code \texttt{AREPO}. Atop the IllustrisTNG galaxy formation model, we implement additional recipes for magnetogenesis in which magnetic energy is injected during supernovae (SNe) and supermassive black hole (SMBH) feedback events, and compare these to simulations initialized with uniform primordial seed fields. Halo magnetic field strengths at $z=0$ are largely similar across seeding models and are primarily amplified and sustained by small-scale and halo-scale dynamo action. Nevertheless, we find differences in magnetic field topology, with SMBH-driven models exhibiting systematically smaller coherence lengths than primordial-only and SNe-only runs. We find that feedback-driven injection accelerates the onset of dynamo growth, leading to more rapid convergence of magnetic field strengths with numerical resolution, particularly in low-mass halos. In the intergalactic medium (IGM), SNe-only injection underproduces magnetic fields relative to inferred lower limits from $\gamma$-ray cascade constraints at both $z=0$ and $z \sim 3$, whereas our specific SMBH-based injection prescription satisfies present-day constraints but remains in mild tension at high redshifts. Reconciling these specific high-$z$ constraints therefore likely requires either modified feedback prescriptions or an additional primordial seeding component.
\end{abstract}

\begin{keywords}
galaxies: halos -- galaxies: magnetic fields -- galaxies: circumgalactic medium -- galaxies: intergalactic medium
\end{keywords}

\maketitle

\section{Introduction}\label{sec:intro}
A multitude of observations suggest that the Universe is threaded by magnetic fields across a remarkable range of cosmic environments. Nearby spiral galaxies host coherent, $\mu$G-strength fields traced by diffuse synchrotron emission and Faraday rotation measurements \citep[][]{fletcher2011,beck2015}, while galaxy clusters exhibit magnetization extending over $\gtrsim$\,100 kpc scales with comparable amplitudes \citep{carilli2002,govoni2004,bonafede2010}. Statistical studies of rotation measures also indicate magnetized plasma in the tenuous gas around galaxies \citep{kronberg2007,vernstrom2019}, both in their circumgalactic medium (CGM) and along sightlines passing through the local intergalactic medium (IGM). Additionally, observations of TeV blazars provide lower limits on the strength of magnetic fields permeating the IGM \citep{neronov2010,tavecchio2011}. 

A broad class of models posit that the origin of these magnetic fields may be traced back to various generation mechanisms in the early Universe, prior to the epoch of galaxy formation. Proposed candidates include magnetogenesis during inflation \citep{turner1988,ratra1992}, cosmological phase transitions such as the electroweak or QCD transitions \citep{vachaspati1991,sigl1997}, and second-order effects associated with cosmological perturbations and recombination-era plasma processes \citep{harrison1970,yamazaki2006}. 

These scenarios typically produce weak seed fields that require subsequent amplification during gravitational collapse and structure formation \citep{widrow2002,marinacci2018}. Turbulent motions generated during halo assembly can drive small-scale dynamo action, leading to rapid growth of magnetic energy \citep{ryu2008,federrath2016}. Primordial models are particularly attractive in that they naturally generate magnetic fields on cosmological scales, potentially accounting for magnetization of the intergalactic medium without invoking large-scale transport from galaxies \citep{durrer2013,subramanian2016}.

An alternative class of models attributes the origin of cosmic magnetism to astrophysical processes associated with galaxy formation. In this picture, magnetic fields are generated locally within galaxies and/or their environments through mechanisms such as the Biermann battery \citep{biermann1950,kulsrud1997}, the Weibel instability \citep{lazar2009}, stellar winds and supernova explosions \citep{rees1987}, or injections by highly magnetized plasma from the accretion disk of supermassive black holes \citep{daly1990,bertone2006}. 

These processes can produce seed fields at significantly higher amplitudes than many primordial scenarios, albeit initially confined to dense regions. Subsequent amplification via small-scale \citep{brandenburg2005,schleicher2010} and large-scale \citep{parker1971,ruzmaikin1988} dynamos can increase magnetic energy within galaxies, while galactic winds and active galactic nuclei (AGN) driven outflows can subsequently transport magnetized plasma into the circumgalactic and intergalactic media \citep{pakmor2020,ramesh2023}.

Despite decades of theoretical development, no single mechanism has yet emerged as a definitive explanation for the observed distribution of magnetic fields across cosmic environments. Primordial models face constraints from the above discussed gamma-ray limits on intergalactic magnetic fields \citep{neronov2010}, while astrophysical scenarios must account for the efficiency and spatial reach of magnetized outflows. Disentangling the relative contributions of early-Universe seeding and later galaxy-driven processes therefore remains an open problem. In particular, quantifying the extent to which stellar feedback and AGN activity can magnetize the CGM and IGM is essential for assessing the relevance, role and importance of astrophysical magnetogenesis \citep[see also][]{garaldi2021}.

The origin of cosmic magnetic fields, however, represents only one aspect of the broader picture. Equally important is their role as a dynamical component in shaping the evolution of baryonic structures across cosmic time.

In galactic environments, magnetic energy densities are often found to be comparable to those of turbulence, thermal gas, and cosmic rays \citep{naab2017}. This near-equipartition implies that magnetic pressure may contribute significantly to the vertical support and force balance of the interstellar medium (ISM) \citep{boulares1990,cox2005}. Through magnetic tension and anisotropic stresses, magnetization alters the propagation of compressive motions and modifies the structure of multiphase gas \citep{elmegreen2004,hennebelle2013}. Magnetized plasmas can further influence disk stability and the development of large-scale instabilities, including spiral structure and magnetorotational turbulence \citep{balbus1991,kim2003,piontek2007}. 
%Even when not energetically dominant, magnetic fields therefore represent a dynamically relevant component of galactic disks.

By contributing magnetic pressure and tension support against gravitational collapse, fields can modify the effective Jeans criterion and influence the fragmentation of molecular clouds, thereby playing a role in regulating star formation within the ISM \citep{mouschovias1976,shu1987}. Observations of Zeeman splitting and dust polarization indicate that many molecular clouds are near critical in their mass-to-flux ratios, suggesting that magnetic support is dynamically significant during cloud evolution \citep{crutcher1999,crutcher2012}. A number of simulations further demonstrate that magnetization can reduce star formation efficiencies, alter filamentary structure, and modify the spectrum of turbulence within collapsing gas \citep{padoan2011,federrath2012}. 

From a theoretical standpoint, turbulent dynamo processes are expected to amplify magnetic energy until back-reaction becomes significant, leading to saturation at a non-negligible fraction of the turbulent kinetic energy \citep{haugen2004,schekochihin2004}. In many galactic environments, this results in plasmas with thermal-to-magnetic pressure ratios $\beta \sim 1$–$10$, placing them in a trans- or super-Alfvénic regime where magnetic stresses can meaningfully alter gas dynamics without fully dominating them \citep{federrath2016}. 

Magnetization is also expected to regulate the transport of non-thermal components, most notably cosmic rays, whose propagation is strongly guided by magnetic field topology. As charged particles spiral along field lines, their transport becomes highly anisotropic, coupling cosmic-ray pressure gradients to the surrounding gas \citep{grenier2015}. This coupling can drive galactic winds, modify disk–halo circulation, and redistribute energy and momentum over $\gtrsim$\,kpc scales \citep{uhlig2012,booth2013,pakmor2016}. The efficiency of cosmic-ray confinement and streaming depends sensitively on field strength and geometry, linking the dynamical impact of cosmic rays directly to the underlying magnetization \citep{skilling1975,farber2018}. 

Magnetization likewise regulates thermal transport in ionized plasmas by rendering conduction highly anisotropic. In weakly collisional environments, heat energy preferentially flows along magnetic field lines, while transverse transport is strongly suppressed \citep{braginskii1965,narayan2001}. This anisotropy can alter the growth of thermal instabilities and modify the evolution of multiphase structure in galactic halos and clusters \citep{sharma2010}. 

Magnetization also introduces qualitatively new plasma instabilities that have no purely hydrodynamic analogues. In stratified atmospheres, anisotropic conduction gives rise to the magnetothermal instability and the heat-flux–driven buoyancy instability, both of which can reorient magnetic field lines and modify convective transport in galaxy clusters and massive halos \citep{balbus2000,parrish2008}. These instabilities influence temperature profiles, regulate conductive heat fluxes, and may ultimately affect the stability of hot atmospheres against cooling \citep{quataert2008}.

On $\sim$\,kpc scales, ordered field structures may influence the interaction between cold clouds and hot ambient media. Draped configurations of fields may contribute to boosting the effective drag force that clouds face, causing them to begin comoving with their ambient media on shorter timescales, thereby reducing the time they are exposed to the wind \citep{dursi2008,mccourt2015,ramesh2024b}. Magnetic pressure support can help clouds counterbalance the thermal pressure of the hot background \citep{nelson2020,fielding2023,ramesh2023b}, while magnetic tension forces may suppress fluid instabilities along their interfaces \citep{ji2018,sparre2020,ramesh2025}. Additionally, even dynamically modest fields can inhibit shear-driven entrainment and alter the cascade of turbulence by introducing anisotropy into the inertial range \citep{goldreich1995,schekochihin2004}. In galactic winds and halo environments, such suppression of mixing may delay the disruption of cold structures and reduce the rate at which metals and thermal energy are homogenized throughout the circumgalactic medium \citep{ji2018,ramesh2026}. 

On larger circumgalactic scales, magnetization can influence properties of halo gas and the structure of outflows \citep{vandevoort2021}. In more massive galaxy clusters, magnetic fields may affect the stability of buoyant bubbles inflated by AGNs, and thus contribute to the regulation of cooling flows \citep{mcnamara2007,ruszkowski2007}. The topology and strength of halo magnetic fields may thus bear directly on the long-term thermodynamic evolution of the circumgalactic and intracluster media.

Given the broad array of physical processes that magnetic fields may influence, understanding their origin and evolution requires following both galaxy formation and magnetic field amplification across a wide range of scales. Cosmological magnetohydrodynamical simulations have therefore become an increasingly important tool for studying cosmic magnetism. To this aim, over the past decade or so, several simulation projects, for instance, but not limited to, Magneticum \citep{dolag2009}, IllustrisTNG \citep{marinacci2018}, FIRE-2 \citep{hopkins2020} and those presented in \cite{vazza2014}, have begun incorporating magnetic fields.

A large fraction of these studies, however, begin from a primordial seed field and focus on its subsequent growth and amplification over cosmic epochs. While this approach has yielded important insights, it leaves open the question of how much of the magnetization observed today may instead originate from astrophysical processes associated with galaxy formation itself. A number of works have taken initial steps in this direction by modelling magnetic field injection from stellar feedback, AGN activity, or processes like the Biermann battery \citep[e.g.][]{vazza2017,garaldi2021}. Yet the impact of these sources remains relatively uncertain, as only a limited subset of the possible galaxy formation models, feedback prescriptions, and magnetic field injection scenarios has been explored to date, leaving much of the relevant parameter space untested.

In this work, we investigate this question by extending the range of astrophysical magnetic field injection models implemented in cosmological simulations. Our aim is to quantify how different sources contribute to the growth and distribution of magnetic fields across cosmic environments, and to assess the extent to which astrophysical seeding can shape the magnetization of galaxies and their surroundings. The rest of this paper is structured as follows: in Section~\ref{sec:methods}, we describe the methods employed in this work, including the new prescriptions used to inject magnetic fields from astrophysical sources. Results are presented and discussed in Section~\ref{sec:results}, and summarized in Section~\ref{sec:summary}.

\section{Methods}\label{sec:methods}
\subsection{Setup overview}
We present results from a suite of large-volume cosmological magnetohydrodynamical simulations (L$_\mathrm{box}$\,$=$\,$25$ Mpc/h) run with the moving-mesh code \texttt{AREPO} \citep{springel2010}. The initial conditions are generated at $z_{\rm{init}}$\,$\sim$\,$127$ with \texttt{MUSIC} \citep{hahn2011} using second-order Lagrangian perturbation theory. For the main body of this work, we sample the initial density field of the above-mentioned volume with $512^3$ pairs of gas cells and dark matter (DM) particles, corresponding to a DM (average baryonic) mass of $\sim$\,10$^7$ (10$^6$)\,M$_\odot$. Following \cite{pillepich2018}, we adopt a Plummer equivalent gravitational softening length of $0.74$\,ckpc for the collisionless components, and a minimum adaptive softening length of $0.19$\,ckpc for gas.

We adopt a cosmology consistent with the Planck 2016 analysis \citep{planck2016}: $h = 0.6774$ [100 km s$^{-1}$ Mpc$^{-1}$], $\Omega_{\rm m} = 0.3089$, $\Omega_{\rm b} = 0.0486$ and $\Omega_\Lambda = 0.6911$.

\subsection{Base model for galaxy physics}\label{ssec:base_physics}
To account for the main physical processes considered important for galaxy formation and evolution, we adopt the widely used IllustrisTNG model \citep[hereafter, TNG model;][]{weinberger2017,pillepich2018}. This includes recipes for radiative cooling of gas, through both primordial \citep{katz1996} and metal-line \citep{wiersma2009} contributions, in the presence of a time-evolving ultra-violet background (UVB) field \citep{fg2009}, formation of stellar particles as parent gas cells cross a specified density threshold \citep{springel2003} and their subsequent evolution, metal enrichment and feedback, the formation of supermassive black holes (SMBHs) and associated feedback \citep{weinberger2017}, among other processes. While we refer the reader to the TNG methods papers \citep{weinberger2017,pillepich2018} for additional details, related to both the physics and their numerical implementations, we here summarize a subset of key points directly relevant to the present work.  

Stellar feedback is realized through a decoupled wind scheme \citep{springel2003}, wherein wind particles stochastically spawn from star-forming gas cells and traverse galactic scales prior to hydrodynamically recoupling, whereby they deposit mass, metals, momentum and energy. In the TNG model, these winds carry both kinetic and thermal energy, with the latter set to $10\%$ of the available wind energy, and is important to avoid spurious star-formation during recoupling \citep{pillepich2018}.

SMBHs are seeded with an initial mass of $\sim 10^{6}$\,$\rm{M_\odot}$ at the potential minima of friends-of-friends halos \citep{davis1985} once they exceed a mass of $\sim$\,$7 \times 10^{10}$\,$\rm{M_\odot}$. Thereafter, SMBHs grow through mergers and gas accretion from their surroundings, while simultaneously releasing feedback energy through one of two modes. At high accretion rates, i.e. when the Bondi-Hoyle accretion rate exceeds a fraction of the Eddington limit (dependent on M$_{\rm BH}$ and additional parameters, see Equation~5 of \citealt{weinberger2017}), SMBHs continuously inject thermal energy into the surrounding gas. At lower accretion rates, feedback instead proceeds through a bursty kinetic mode wherein momentum is injected into the ambient medium. In the TNG model, the transition between these two modes typically occurs around M$_{\rm BH}$\,$\sim$\,$10^8$\,$\rm{M_\odot}$, roughly corresponding to a halo mass of M$_{\rm{200c}}$\,$\sim$\,$10^{12}$\,M$_\odot$ \citep{pillepich2021}. In both cases, the energy deposition is implemented through a kernel-weighted prescription over a specified number of neighboring gas cells around the SMBH \citep{weinberger2017}.

Lastly, the TNG model also follows the growth, evolution and dynamical impact of magnetic fields by solving the coupled equations of ideal magnetohydrodynamics \citep[MHD;][]{pakmor2013}: to the best of our knowledge, in all simulations that employ the TNG model, a uniform field is seeded with a value of $\sim$\,$10^{-14}$\,cG along an arbitrary direction in the initial conditions ($z_{\rm{init}}$\,$\sim$\,$127$), and the \cite{powell1999} eight-wave scheme is used to control divergence errors such that $\nabla \dcdot \vec{B}$\,$\sim$\,$0$.

\subsection{Additional magnetic field seeding mechanisms}
In order to explore the possible impact of a subset of galactic magnetogenesis processes, we implement two additional models. As described more extensively in their respective sub-sections below, these are aimed at mimicking, in a sub-grid fashion, the galactic-scale impact of possible injections during supernovae and SMBH feedback events. In both cases, we load a given amount of magnetic energy ($\rm{E_{B}}$) into a wind particle at the time of spawning, and thereafter inject a magnetic dipole of strength \citep{beck2013}

\begin{equation}
    m = \sqrt{ 
              \frac{3 r^3_{\rm{soft}} \left(r^3_{\rm{soft}} + r^3_{\rm{inj}} \right)
              \rm{E_{B}}}
              {4 \pi r^3_{\rm{inj}}}
        } 
\label{eq:dipole_str}
\end{equation}

oriented along a random direction into the nearest $64 \pm 4$ gas cells when the wind hydrodynamically recouples. $r_{\rm{inj}}$ is the radius that encloses the above selection of gas cells, $r_{\rm{soft}}$ corresponds to a softening scale used to avoid discontinuities when the site of injection is close to the position of the gas cell, and the combination of factors in Equation~\ref{eq:dipole_str} normalizes the dipole such that the total magnetic energy injected into the truncated region bounded by $r_{\rm{soft}}$ and $r_{\rm{inj}}$ is conserved \citep[see also][]{donnert2009}. If multiple dipoles are simultaneously injected in a given region, they are linearly summed.

\subsubsection{SN-driven winds}\label{sssec:sn_winds}
Supernovae are a plausible source of magnetic field generation in galaxies, as magnetic fields can be generated and amplified within supernova remnants through plasma processes and turbulence driven by the explosion \citep{bell2004,hanayama2005}. These locally generated fields may thereafter be advected into the surrounding interstellar medium by the ejecta, where they can subsequently mix, spread and undergo further amplification through turbulence and galactic-scale dynamo action \citep{xu2017,rincon2019}.

Motivated by this picture, when a wind particle associated with stellar feedback is spawned (Section~\ref{ssec:base_physics}), we load a fraction $\tau_{\rm{mag,SN}}$ of the total wind energy in magnetic form. The available kinetic fraction is thus reduced to $1 - \tau_{\rm{mag,SN}} - \tau_{\rm{thm,SN}}$, where the thermal component of the wind is left fixed at $\tau_{\rm{thm,SN}}$\,$=$\,$10\%$. All other technical details regarding their spawning and recoupling also remain unchanged \citep{pillepich2018}.

\subsubsection{SMBH-driven winds}\label{sssec:bh_winds}
SMBH-related activity, in particular in the high accretion regime, is commonly associated with fast, ionized disc winds observed as, e.g.,  broad absorption line outflows, which trace powerful mass-loaded ejections from the accretion flow \citep{weymann1991,elvis2000}. These winds are launched from the inner regions of the accretion disc, where magnetized, differentially rotating plasma is theoretically expected, and where magnetic fields can be sustained and amplified through magneto-rotational instability and shear-driven dynamo action \citep{balbus1991,hawley2015}. As the winds are accelerated to high velocities and propagate away from the nucleus, the magnetized plasma is transported into the host galaxy, providing a channel through which magnetic energy associated with SMBH feedback can be deposited on galactic scales \citep[see also][]{konigl1994}.

Similar to Section~\ref{sssec:sn_winds}, we numerically implement this galactic-scale transport using wind particles \citep[see also][]{su2021}. First, we modify the high-accretion quasar mode of SMBH feedback (Section~\ref{ssec:base_physics}) as follows: in the TNG model, the high-accretion channel proceeds exclusively with a continuous thermal energy dump of 

\begin{equation}
    \Delta\dot{E}_{\rm{high}} = \epsilon_{\rm{f,high}} \epsilon_{\rm{r}} \dot{M}_{\rm{BH}} c^2
\label{eq:edot_high}
\end{equation}

for every $\dot{M}_{\rm{BH}}$ of mass accreted onto the SMBH \citep{weinberger2017}. Here, $\epsilon_{\rm{r}}$ is the radiative efficiency, and $\epsilon_{\rm{f,high}}$ corresponds to the fraction of energy that couples to the ambient gas, which in the TNG model are set to $0.2$ and $0.1$, respectively (see Table~1 of \citealt{pillepich2018}). 

In our new implementation, for SMBHs in the high-accretion mode, the energy deposition is no longer solely thermal in nature, but rather this only accounts for a fraction $\tau_{\rm{thm,BH}}$ of the energy computed in Equation~\ref{eq:edot_high}. The remainder is portioned into kinetic and magnetic components, parameterized by fractions $\tau_{\rm{kin,BH}}$ and $\tau_{\rm{mag,BH}}$, respectively. To avoid a total re-calibration of the TNG model, we keep the numerics of the thermal dump unchanged, i.e. this continues to take place by a kernel-weighted, continuous injection into a specified number of gas cells around the SMBH. 

For the kinetic and magnetic components, we implement a new `massless wind', i.e. numerical carrier particles that only deposit momentum and a magnetic dipole upon recoupling, and no mass, which we spawn whenever the energy accumulated in the kinetic channel exceeds a threshold $E_{\rm{min,KE}}$ \citep[see also][]{sullivan2026}. Inspired by the large-scale impact of SMBH-feedback in the SIMBA simulations \citep{dave2019}, we parameterize an outflow velocity for these wind particles based on the mass of the SMBH ($\rm{M_{BH}}$) as 

\begin{equation}
    v_{\rm{wind}} [\rm{km/s}] = 500 + \frac{500}{3} \rm{log_{10}\left(\frac{M_{BH}}{10^6 M_\odot}\right)}
\end{equation}

along a random direction, since averaged over multiple spawned wind particles, the transport is assumed to be isotropic at the injection site. In an attempt to ensure that the momentum and magnetic dipole deposited into a given gas cell upon recoupling is roughly independent of numerical resolution, we assign a momentum of $m_{\rm{bar}} v_{\rm{wind}}$ per wind, where $m_{\rm{bar}}$ is the average baryonic mass resolution. Equivalently, the threshold $E_{\rm{min,KE}}$ is $\frac{1}{2} m_{\rm{bar}} v_{\rm{wind}}^2$, and the magnetic energy loaded into the wind is thereby scaled as $\rm{E_{B}}$\,=\,$\frac{E_{\rm{BH,mag}}}{E_{\rm{BH,kin}}}$ \,$\left( \frac{1}{2} m_{\rm{bar}} v_{\rm{wind}}^2 \right)$, where $E_{\rm{BH,kin}}$ ($E_{\rm{BH,mag}}$) is the energy accumulated in the kinetic (magnetic) channel. 

If $E_{\rm{BH,kin}}$ exceeds $2 E_{\rm{min,KE}}$, we simultaneously spawn $\left\lfloor \frac{E_{\rm{SMBH,kin}}}{E_{\rm{min,KE}}} \right\rfloor$ wind particles. The conditions for wind recoupling are identical to those spawned during stellar feedback, as described in \cite{pillepich2018}. Note that, in this work, we do not modify the low-accretion SMBH feedback channel of the TNG model.

\begin{figure*}
    \centering
    \includegraphics[width=18cm]{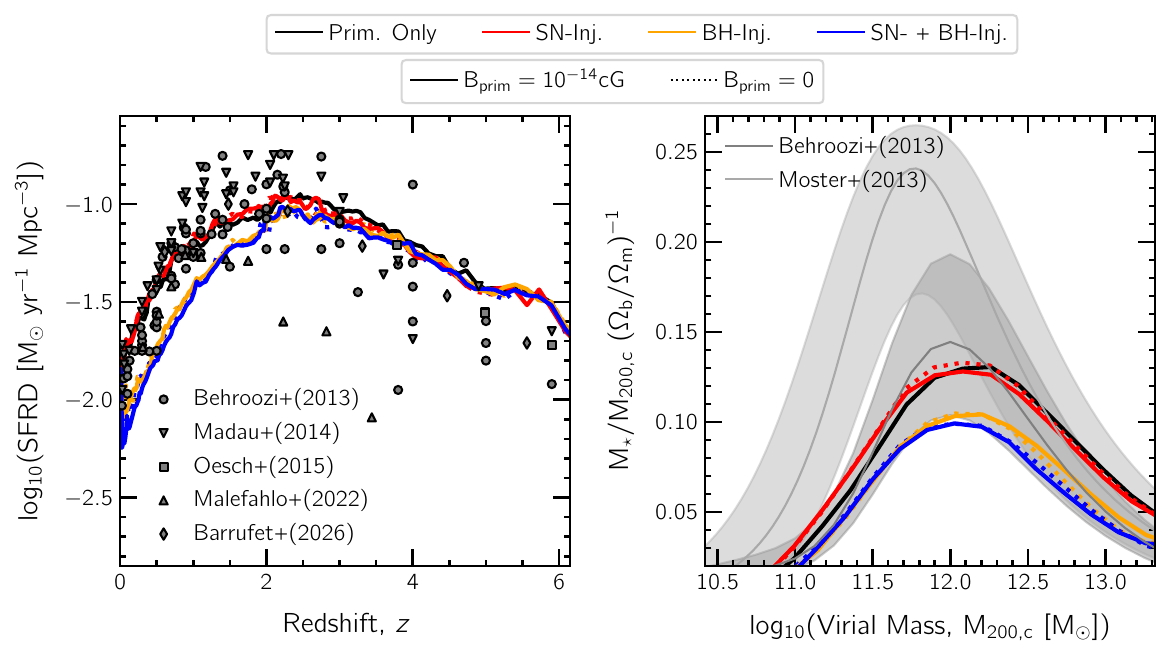}
    \caption{\textbf{Calibration of galaxy population properties across the simulation suite:} Left: cosmic star-formation rate density evolution as a function of redshift compared against observational constraints from \protect{\cite{behroozi2013,madau2014,oesch2015,malefahlo2022,barrufet2026}}. Right: stellar-to-halo mass relation at $z$\,$=$\,$0$ for central galaxies, compared to empirical estimates from \protect{\cite{behroozi2013,moster2013}}. Different colors indicate the various magnetic seeding prescriptions, while solid and dotted lines distinguish runs initialized with primordial seed fields and zero primordial field, respectively. The free parameters of the SN- and BH-driven magnetogenesis models are chosen to maximize the impact of magnetic energy injection while maintaining reasonably good agreement with key observational constraints on the galaxy population.}
    \label{fig:int_prop}
\end{figure*}

\begin{figure*}
    \centering
    \includegraphics[width=18cm]{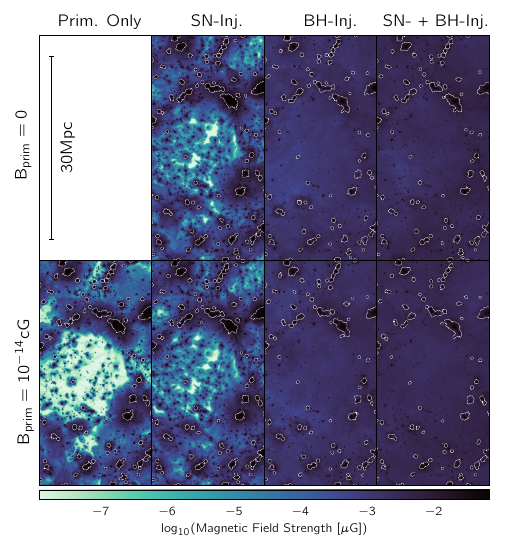}
    \caption{\textbf{Projected $\mathbf{z=0}$ magnetic field strength across a representative cosmological volume for different seeding models:} Mass-weighted magnetic field strengths are projected through the full 25\,Mpc/h simulation volume for runs initialized either with no primordial magnetic field (top row) or with a uniform primordial seed field of $\sim$\,$10^{-14}$\,cG (bottom row). Columns correspond to the primordial-only, SN-driven, BH-driven, and combined SN+BH injection models. White contours indicate mark the projected extent of halos whose radial magnetic field strength profiles are numerically converged at the fiducial resolution in the primordial-only run (bottom-left), as demonstrated in Figure~\ref{fig:app_res_conv_rad_prof}. Within these halos, the resultant magnetic field strengths are broadly similar across all seeding prescriptions, while differences emerge in smaller systems and the diffuse environments surrounding halos. The latter reflect genuine differences in large-scale magnetization, although residual numerical convergence effects may also contribute.}
    \label{fig:main_intro_vis}
\end{figure*}

\section{Results}\label{sec:results}

\subsection{Calibration of wind parameters}\label{ssec:wind_calib}

As described in Section~\ref{sec:methods}, our new implementation of the wind-based magnetic field transport is parameterized by three quantities: a magnetic energy fraction for SN-driven winds ($\tau_{\rm mag,SN}$), and magnetic and kinetic energy fractions for SMBH-driven winds ($\tau_{\rm mag,BH}$ and $\tau_{\rm kin,BH}$). While these are ultimately linked to the efficiency with which magnetic fields are generated and transported by relevant astrophysical processes, robust theoretical or observational constraints on their values remain scarce. We therefore treat them as free parameters and adopt the largest values that do not substantially degrade the agreement between the simulated galaxy population and a subset of observational constraints. In this sense, our goal is not to calibrate a first-principles model of magnetic field generation, but rather to maximize the potential impact of astrophysical magnetogenesis while maintaining a galaxy population that remains broadly consistent with observations.

In particular, we calibrate against available observational data of the evolution of the cosmic star formation rate density (SFRD), and the stellar-to-halo mass relation at $z=0$, as shown in the left and right panels of Figure~\ref{fig:int_prop}, respectively. In both panels, we color curves by the different magnetogenesis models considered: black corresponds to a primordial-only seed field, implemented by assigning a uniform field of $\rm{B_{prim}}$\,$=$\,$10^{-14}$\,cG along the $z$-axis of the box in the initial conditions ($z$\,$\sim$\,$127$). The red curve includes injection by SN-winds with $\tau_{\rm mag, SN}$ set to $1\%$, the orange curve to the inclusion of SMBH-winds with $\tau_{\rm kin, BH}$\,$=$\,$0.1\%$ and $\tau_{\rm mag,BH}$\,$=$\,$0.02\%$, and the blue to a variation in which both SN- and SMBH-winds operate, with the same parameter choices as above. 

Solid curves correspond to those where $\rm{B_{prim}}$\,$=$\,$10^{-14}$\,cG, as described above, while $\rm{B_{prim}}$ is set to zero in the dotted curves. The red-, orange- and blue-dotted curves thus correspond to runs where the origin of fields is `astrophysical-only', the solid black to a `primordial-only' case, while the remainder of lines are hybrids of the above two cases. Note that we do not consider the case with no astrophysical injection and $\rm{B_{prim}}$\,$=$\,$0$, i.e. one in which no magnetic fields exist throughout the simulation (see \citealt{pillepich2018} for a discussion on the impact of this variation on galaxy properties within the scope of the TNG model).

In the left panel, we include available constraints from \cite{behroozi2013,madau2014,oesch2015, malefahlo2022, barrufet2026}, shown by the different scatter points. At epochs prior to cosmic noon ($z \gtrsim 2$), the evolutionary profiles between the different magnetogenesis runs are largely similar, and in broad agreement with the various data points. All the curves peak at $z \sim 2$ and begin to turn over towards lower-$z$, albeit with a sharper decline in the runs including the SMBH-winds: while the red and black curves are in excellent agreement with the considered constraints, the orange and blue are slightly offset vertically (see also Section~\ref{ssec:disc} and Appendix~\ref{ssec:kin_vs_mag_inj}), although still intersecting some of the low-$z$ points of \cite{behroozi2013}.

In the panel on the right, in the light and dark shaded bands, we show empirical constraints from \cite{moster2013} and \cite{behroozi2013}, respectively. Akin to earlier, the red curves closely follow the black, while the orange and blue lines are vertically offset to lower M$_\star$ at fixed M$_{\rm{200c}}$, a direct result of reduced star formation rates discussed above. Furthermore, in both panels, note that the dotted and solid curves of a given astrophysical injection model do not vary significantly, i.e. the presence (or absence) of a primordial seed field does not have a dominant impact on the two integrated properties explored here.

While not shown explicitly, based on various tests performed, we mention that increasing $\tau_{\rm mag, SN}$ beyond the current setting ($1\%$) leads to SFRD values that exceed the observational constraints. We find that this is due to the kinetic component of the wind energy being reduced in proportion, thus leading to a weaker regulation of star formation. On the contrary, increasing $\tau_{\rm kin, BH}$ leads to a decrease in SFRD, as the kinetic energy dumps begin driving gas out of galaxies and into the halo, effectively reducing the amount of fuel available for sustained star formation. Increasing $\tau_{\rm mag, BH}$ leads to elevated SFRD values at low-$z$, due to indirect couplings that modify the mass growth profiles of SMBHs over time, thus delaying the onset of the kinetic mode of SMBH feedback in the TNG model that plays an important role in quenching high-mass galaxies \citep[e.g.][]{zinger2020,ayromlou2023}. 

While a certain degree of fine-tuning of other relevant numerical parameters may make it possible to further vary $\tau_{\rm mag,SN}$, $\tau_{\rm kin,BH}$, and $\tau_{\rm mag,BH}$ while still retaining a realistic galaxy population, we do not pursue such an extensive re-tuning of the model in the present work, as it would require a pervasive analysis of the model predictions. Instead, we adopt the above values as fiducial, near-maximal configurations and investigate the extent to which they modify the growth and distribution of magnetic fields across different cosmic environments. %This allow us to rely on the vast body of literature comparing the IllustrisTNG model to a variety of observations.

To motivate the analyses that we present through the rest of the paper, in Figure~\ref{fig:main_intro_vis}, we show mass-weighted projections of magnetic field strengths through a rectangular [25, 12.5, 25]\,Mpc/h volume at $z$\,$=$\,$0$, projected along one of the long-edges. From left to right, columns correspond to the primordial-only, SN-driven, BH-driven, and combined SN+BH injection models. In the top row, we show models with no primordial seed field, while we show runs with $\rm{B_{prim}}$\,$=$\,$10^{-14}$\,cG in the bottom row. Similar to Figure~\ref{fig:int_prop}, the top row can thus be categorized as those variations wherein the origin of the field is `astrophysical-only', the bottom-left as a `primordial-only' origin, and the remainder of panels as hybrid cases.

White contours overlaid in the foreground correspond to halos which satisfy a mass threshold of M$_{\rm{200c}}$\,$\gtrsim$\,$10^{11.5}$\,M$_\odot$; these are ones wherein the radial magnetic field strength profiles within R$_{\rm{200c}}$ are numerically converged at the fiducial resolution in the primordial-only run (bottom-left), as demonstrated in Figure~\ref{fig:app_res_conv_rad_prof}. Within these systems, the field strengths visually appear broadly similar across seeding models. We show below that this also holds quantitatively. Outside these structures, i.e. within smaller halos and in the volume permeating the space in between halos, clear differences emerge. These may arise from differences in large-scale magnetization between models, residual numerical convergence effects, or both, and we dissect their relative contributions in detail throughout the remainder of this work. 

\begin{figure*}
    \centering
    \includegraphics[width=18cm]{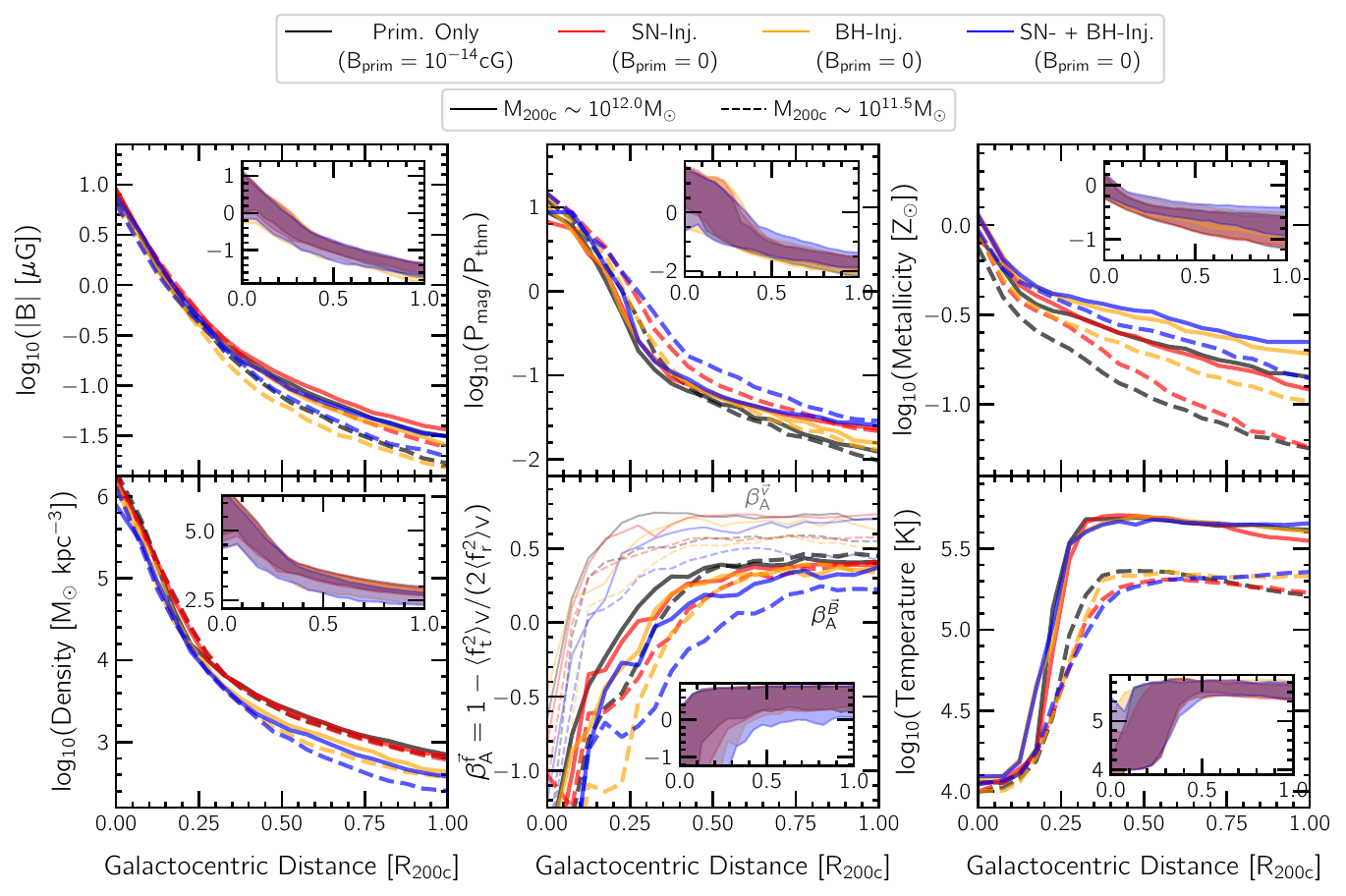}
    \caption{\textbf{Radial thermodynamic and magnetic profiles of halos at $\mathbf{z=0}$:} Median, spherically-averaged radial profiles are shown for halos of mass M$_{\rm{200c}}$\,$\sim$\,$10^{12}$\,M$_\odot$ (solid lines) and $\sim$\,$10^{11.5}$\,M$_\odot$ (dashed lines). Panels show, in clockwise order starting from the top-left: magnetic field strength, magnetic-to-thermal pressure ratio, metallicity, temperature, anisotropy of field lines, and gas density, as functions of galactocentric distance normalized by R$_{\rm{200c}}$. Through the various shaded bands, insets show the 1$\sigma$-variation of M$_{\rm{200c}}$\,$\sim$\,$10^{12}$\,M$_\odot$ halos in the different runs. Across all seeding prescriptions, halo magnetic field strengths at $z$\,$=$\,$0$ converge to broadly similar values, indicating that the final halo magnetization is largely insensitive to the initial seeding mechanism. In contrast, the  profiles of the other physical quantities explored here differ more noticeably in a subset of other suites, for instance the metallicity profiles in the BH-driven models due to the additional kinetic energy injection associated with BH winds, although these variations do not appear to strongly affect the resulting halo magnetic field strengths.}
    \label{fig:radial_prof}
\end{figure*}

\begin{figure*}
    \centering
    \includegraphics[width=18cm]{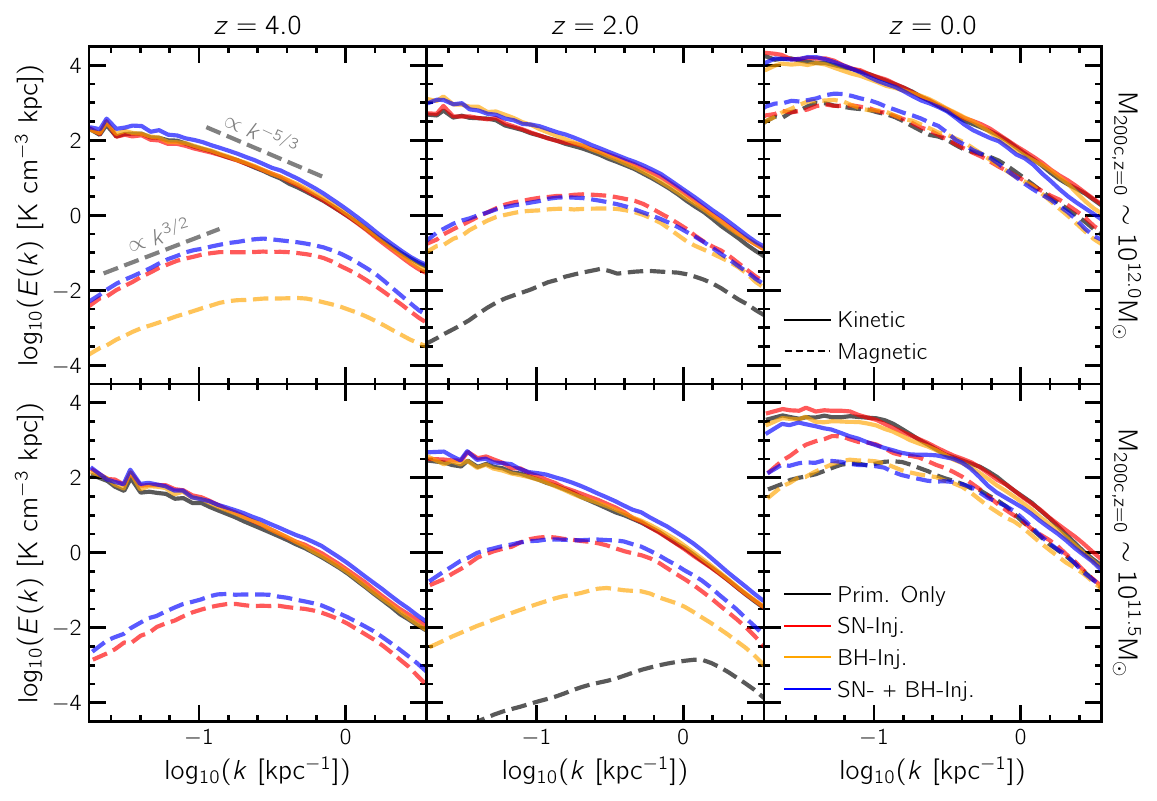}
    \caption{\textbf{Evolution of kinetic and magnetic power spectra within halos:} From left to right, power spectra are shown at $z$\,$=$\,$4$, $2$, and $0$ for halos with present-day masses M$_{\rm{200c}}$\,$\sim$\,$10^{12}$\,M$_\odot$ (top row) and M$_{\rm{200c}}$\,$\sim$\,$10^{11.5}$\,M$_\odot$ (bottom row). Solid and dashed lines denote kinetic and magnetic spectra, respectively, while colors correspond to different seeding mechanisms. Grey dashed lines indicate the slopes expected from the Kazantsev ($k^{3/2}$) and Kolmogorov ($k^{-5/3}$) theories. Feedback-driven magnetic injection accelerates the onset of small-scale dynamo amplification, allowing magnetic power to build up more rapidly on resolved scales and thereby leading to faster numerical convergence of halo magnetic field strengths, particularly in lower-mass systems (see also Figure~\ref{fig:app_res_conv_rad_prof}).}
    \label{fig:power_spectra}
\end{figure*}

\begin{figure*}
    \centering
    \includegraphics[width=18cm]{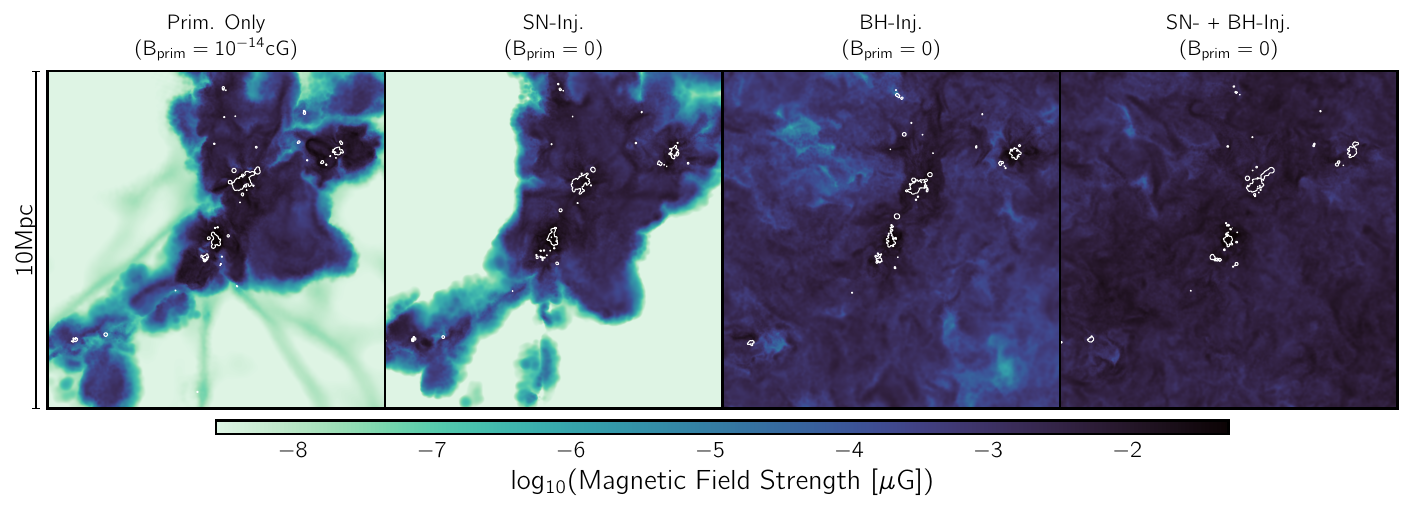}
    \caption{\textbf{Large-scale magnetic field structure around a Milky Way-mass halo:} Projected magnetic field strength maps at $z$\,$=$\,$0$ are shown for a region centered on a M$_{\rm{200c}}$\,$\sim$\,$10^{12}$\,M$_\odot$ halo, cross-matched across the different seeding models. Each panel spans $\pm$\,$5$\,Mpc in the image plane and a $\pm$\,$100$\,kpc slice along the projection direction, and white contours in the foreground show regions of non-zero stellar density in the same thin slice. In the primordial-only case (left panel), filamentary magnetic structures are visible, reflecting the compression and amplification of initially volume-filling seed fields during large-scale structure formation. These filamentary features are largely absent in the SN-only model (center-left panel), where magnetic fields are injected locally by galactic winds rather than being frozen into the collapsing cosmic web. In the BH-driven models, large-scale magnetic pollution from feedback spreads over the surrounding environment, obscuring the contrast seen between primordial filamentary magnetization and locally injected SN-driven fields in the preceding panels.}
    \label{fig:filament_vis}
\end{figure*}

\subsection{Properties of gas within halos}\label{ssec:halo_gas}
In Figure~\ref{fig:radial_prof}, we begin with a quantitative analysis of six different physical properties of gas within halos at $z$\,$=$\,$0$, focusing on spherically-averaged radial profiles. In the main panels, curves are colored by the seeding prescription(s) considered. The black corresponds to a `primordial-only' run, where we begin with a uniform field of $\rm{B_{prim}}$\,$=$\,$10^{-14}$\,cG, as described above. The other three correspond to `astrophysical-only' runs, where $\rm{B_{prim}}$ is set to zero. In the red (orange) line, we include contributions only from SN- (BH-) winds, while the blue is the more generic case where magnetic dipole injections from both winds are active. Although not shown explicitly, we mention that the halo gas properties explored here are largely insensitive to the presence of an initial primordial seed field once alternate astrophysical prescriptions are activated\footnote{Furthermore, unless explicitly mentioned, through the rest of the draft, we only contrast the `primordial-only' run versus the `astrophysical-only' variations, i.e. we do not explicitly consider the hybrid cases.}, a trend also visible qualitatively in Figure~\ref{fig:main_intro_vis}.

We consider two different mass bins of width $\pm$\,$0.2$\,dex: M$_{\rm{200c}}$\,$\sim$\,$10^{12}$\,M$_\odot$ shown in solid curves, and M$_{\rm{200c}}$\,$\sim$\,$10^{11.5}$\,M$_\odot$ in dashed. The motivation for this choice is twofold: (i) the field strengths of halos less massive than M$_{\rm{200c}}$\,$\sim$\,$10^{11.5}$\,M$_\odot$ are not numerically converged in the primordial-only run (Appendix~\ref{sec:app_res_conv}), and (ii) given the relatively small simulation domain (L$_{\rm{box}}$\,$\sim$\,25\,Mpc/h), halos more massive than M$_{\rm{200c}}$\,$\sim$\,$10^{12}$\,M$_\odot$ are relatively few in number. Moreover, note that the SMBHs in halos above this threshold are preferentially in the kinetic mode \citep[e.g.][]{pillepich2024}, where we do not include a magnetic channel in the present work. Lastly, in the insets, to provide an idea of variation across the sample, we show 1$\sigma$-bands of M$_{\rm{200c}}$\,$\sim$\,$10^{12}$\,M$_\odot$ halos from the different runs.

In the top-left panel, we begin by exploring radial profiles of magnetic field strengths. At the Milky Way-mass scale (M$_{\rm{200c}}$\,$\sim$\,$10^{12}$\,M$_\odot$), median profiles are largely similar across all seeding models. The same holds for halos in the less massive M$_{\rm{200c}}$\,$\sim$\,$10^{11.5}$\,M$_\odot$ bin at distances $\lesssim$\,$0.25$\,R$_{\rm{200c}}$, with the red dashed curve exhibiting marginally elevated levels of magnetization at larger distances. The inset shows that the halo-to-halo diversity is typically $\sim$\,$0.6-1.0$\,dex, depending on distance, across all seeding models.

We next consider the density of gas in the bottom-left panel. While profiles are largely similar between the red and black curves, at both halo masses, differences begin to arise in the runs with the BH-wind seeding active, particularly away from the central galaxy ($\gtrsim$\,$0.25$\,$R_{\rm{200c}}$). For instance, at the virial radius, the blue solid (dashed) line is vertically offset by $\sim$\,0.2 (0.4) dex with respect to the corresponding black curves. As we discuss further below, we understand this to be the result of large-scale shock heating triggered by rapid outflows driven by the BH wind.

An important quantity that (partly) characterizes the dynamical impact of fields is the ratio of magnetic to thermal pressures, which we show in the top-center panel.  Although magnetic field profiles are largely similar across runs (top-left panel), more pronounced differences are noticeable in the nature of pressure support. For example, in both black curves, the ratio is as large as $\sim$\,$10$ at the very center of the halo, and gradually decays to $\sim$\,$0.01$ at the virial radius. In the runs with SN-winds seeding active, the drop is less steep, with values higher by a factor of $\sim$\,$2-3$ at R$_{\rm{200c}}$. The most extreme difference is present in the blue dashed curve, broadly consistent with the reduced gas densities, leading to lower thermal pressures. The orange curves, i.e. with only BH-wind active, are intermediate between the two cases, in both mass bins.

Expanding beyond the magnitude of the magnetic field, in the bottom-center panel, we aim to quantify the local morphology of field lines via the anisotropy parameter $\beta_{\rm{A}}$. Following \cite{lehle2026}, we define this for any vector field $\vec{\rm f}$ as

\begin{equation}
    \beta_{\rm{A}}^{\vec{\rm f}} = 1 - \frac{\langle \rm{f_{t}^2} \rangle_{\rm{V}}}{2 \langle \rm{f_{r}^2} \rangle_{\rm{V}}}
    \label{eq:beta_A}
\end{equation}

where $\langle \rangle_{\rm{V}}$ corresponds to volume-weighted quantities, and $\rm{f_{t}}$ and $\rm{f_{r}}$ to the tangential and radial components of the field, respectively. A value of $1$ thus quantifies a field that is radial, $\sim$\,$0$ to one that is roughly isotropic, and $<$\,$0$ to tangentially oriented fields \citep[see also][]{binney1987}. 

Across all seeding models, magnetic field lines are largely tangential in the inner halo ($\lesssim$\,$0.1-0.15$\,R$_{\rm{200c}}$), qualitatively consistent with the presence of ordered field lines along the spiral arms of star-forming, disky galaxies \citep{pakmor2014}. Towards larger distances, the fields transition to more isotropic topologies, before turning preferentially radial closer to the virial radius, qualitatively similar to X-shaped magnetic structures observed in the halos of nearby edge-on galaxies \citep[e.g.][]{terral2017,krause2020}. At a given halo mass, the above `switch' in configuration is typically more gradual in the runs with wind-driven magnetic injection active: as an example, the solid black curve rises sharply to cross $\beta_{\rm{A}}^{\vec{\rm B}}$\,$=$\,$0$ at $\sim$\,$0.25$\,R$_{\rm{200c}}$, while this flip is horizontally offset to $\sim$\,$0.4$\,R$_{\rm{200c}}$ in the solid blue curve.

For reference, the thinner curves in the background additionally show analogous profiles for the velocity field ($\beta_{\rm{A}}^{\vec{\rm v}}$). Interestingly, between any given set of curves, i.e. same halo mass and seeding prescription, trends are qualitatively similar. We therefore posit that the differences in the anisotropy of the magnetic field are primarily driven by differences in the anisotropy of the corresponding velocity field. 

We next examine the radial distribution of metallicity (top-right panel). At the very center of the halo, i.e. within the galaxy, the metallicity is primarily a function of halo mass, and rather invariant across seeding models, consistent with the expectation of a tight stellar mass to gas-phase metallicity relation \citep{tremonti2004,torrey2019}. Transitioning to larger distances, differences begin to emerge between the different runs. In particular, in the variations which include the BH-wind, metallicities of halo gas are significantly higher, a direct result of the associated kinetic energy injection driving metal-enriched gas away from galaxies. Although such outflows are also expected to transport highly magnetized gas \citep[e.g.][]{pakmor2020,ramesh2023}, no significant boost in field strengths is seen in the top-left panel, suggesting that the transport of magnetized gas alone is insufficient to maintain substantially enhanced magnetic field strengths within halos (see also the discussion of Figure~\ref{fig:power_spectra}). 

Lastly, in the bottom-right panel, we conclude the exploration of halo gas properties by examining the radial trend of temperature. To first order, the shapes and normalizations of the associated profiles are set by gravity: curves begin at T\,$\sim$\,$10^4$\,K in the inner halo, dominated by cool galactic gas, increase to a characteristic virial temperature which depends on halo mass, and begin gradually reducing towards larger distances. Interestingly, in the runs with BH-winds active, the temperatures begin to increase modestly at galactocentric distances close to the virial radius, due to large-scale shock heating triggered by the rapid BH-wind driven outflows. This is reminiscent of the temperature structure seen in the SIMBA simulation \citep{dave2019}, upon which the BH-wind launch velocity is based.

Taken together, these profiles indicate that the thermodynamic state of halo gas can retain a clear imprint of the underlying feedback model, particularly when BH-wind driven outflows are present. The magnetic field strengths within halos, however, remain remarkably similar across the adopted seeding prescriptions \citep[see also][]{garaldi2021}. Differences in field topology are more apparent, but appear to largely trace corresponding changes in the velocity structure of the halo gas rather than the seeding mechanism alone. This raises the question of why substantially different magnetogenesis channels ultimately produce such similar magnetic field strengths. 

To address this, in Figure~\ref{fig:power_spectra}, we examine magnetic and kinetic energy power spectra of halo gas, providing insight into how different seeding channels populate spatial scales and how subsequent amplification processes reshape the distribution of energy over cosmic time. As before, at $z$\,$=$\,$0$, we select halos at two distinct mass scales: M$_{\rm{200c}}$\,$\sim$\,$10^{12}$\,M$_\odot$ shown in the top row, and $\sim$\,$10^{11.5}$\,M$_\odot$ in the bottom. 

In particular, we select all systems within a narrow $\pm$\,$0.025$\,dex bin centered on the corresponding halo mass. Following this, we track the sample back in time with merger trees constructed using \texttt{SubLink} \citep{rg2015,nelson2015}, and show the median curves at $z$\,$\sim$\,$4$, $2$, and $0$, in the columns from left to right, respectively. Largely similar to \cite{pakmor2020}, the power spectra are obtained by computing the absolute square of the Fourier transforms of $\vec{B}\,/\sqrt{8 \pi}$ and $\sqrt{\rho} \vec{v}$ over a selection of gas cells within a constant physical radial limit defined by the median virial radius of the corresponding bin at $z=0$ (R$_{\rm{200c, z=0}}$), placed within a zero-padded box of size $4$\,R$_{\rm{200c, z=0}}$. In all panels, curves are colored according to the relevant seeding prescription(s), with the color scheme consistent with that of Figure~\ref{fig:radial_prof}; solid (dashed) lines correspond to kinetic (magnetic) spectra.

In all panels shown here, i.e. across both mass bins and three cosmic epochs, kinetic energy spectra are largely consistent across seeding models: all exhibit a slope similar to a Kolmogorov-like scaling (\citealt{kolomogorov1941}; $\propto k^{-5/3}$, shown in the top-left panel for reference through a gray dashed line), albeit with a time-evolving normalization. That is, although a turbulent cascade is under operation at epochs as early as $z$\,$\sim$\,$4$ at both these halo mass scales, the corresponding turbulent velocities evolve over time. Interestingly, despite an additional kinetic energy injection in the runs with the BH-wind active, no systematic boost is present in the kinetic spectra, suggesting that only a small fraction of this energy is retained in the form of turbulent motions within the halo.

The magnetic energy spectra, however, are qualitatively different. First, in both mass bins at early cosmic epochs ($z$\,$\gtrsim$\,$2$), a clear signature of ongoing dynamo action is visible, with power on small scales approaching saturation while the larger-scale component remains consistent with a Kazantsev-like scaling (\citealt{kazantsev1968}; $\propto k^{3/2}$), as expected for turbulent dynamo amplification. By $z$\,$=$\,$0$, the magnetic energy spectra saturate at amplitudes corresponding to $O(1-10\%)$ of the kinetic energy \citep[see also the detailed discussion in][]{pakmor2020}, and are largely similar across all runs, suggesting that the eventual magnetic energy budget is regulated primarily by a halo-scale dynamo rather than the details of the implemented seeding prescription.

Second, although magnetic power spectra are broadly similar across models at $z$\,$=$\,$0$, clear differences exist at earlier epochs. In particular, the runs with magnetic injection via SN-winds (red and blue curves) have significantly higher magnetic power across scales, and the approach to saturation is effectively accelerated. Injection via BH-driven winds alone is unable to match the above rate of rapid saturation, at least within the framework of the TNG model that we employ, i.e. if SMBHs were to be seeded earlier, in less massive halos where SN-driven feedback dominates, the above trend may be altered. In any case, the saturation of the magnetic energy budget is the least rapid in the primordial-only run, as this relies exclusively on magnetized gas to be transported from the galaxy into the halo \citep{pakmor2020}, as opposed to explicitly depositing magnetic energy as in the other astrophysical seeding runs.

Lastly, we discuss the trend with halo mass. The approach to saturation generally proceeds more rapidly in more massive systems, reflecting the earlier onset of star formation, feedback activity, and turbulent amplification within their progenitors. The above trend is also strongly affected by numerical resolution, as this directly determines when halos become sufficiently resolved to sustain efficient dynamo amplification (\citealt{pakmor2024}; see also Appendix~\ref{sec:app_res_conv}). Viewed in this context, Figure~\ref{fig:power_spectra} demonstrates that our astrophysical seeding prescriptions accelerate the growth of magnetic energy within halos and thereby promote more rapid convergence of halo magnetic field strengths with numerical resolution. This effect is particularly pronounced in lower-mass systems (see also Figure~\ref{fig:main_intro_vis}, and associated discussion), which often fail to achieve convergence at the resolution levels currently accessible to large-volume cosmological magnetohydrodynamical simulations \citep[][]{pakmor2024}.

\subsection{Properties of gas along filamentary structures}\label{ssec:fil_gas}
Having explored the impact of different seeding models on gas within collapsed halos, we now shift our focus to the diffuse gas outside such structures. To set the stage, in Figure~\ref{fig:filament_vis}, we begin with a visualization of projected magnetic field strength at $z$\,$=$\,$0$ around a Milky Way-mass system (M$_{\rm{200c}}$\,$\sim$\,$10^{12}$\,M$_\odot$). To be consistent across runs, we select a representative example in the primordial-only box (first panel from the left), and cross-match to identify the corresponding analog in the other variations, shown in the second to fourth panels, for the SN-only wind, BH-only wind, and the joint SN- and BH-winds, respectively. Each panel spans $\pm$\,$5$\,Mpc in the image plane and a thin $\pm$\,$100$\,kpc slice along the projection direction, and white contours in the foreground mark regions of non-zero stellar density within the same slice.

The visual differences between the seeding models are readily apparent in the gas surrounding the selected halo. In the primordial-only run, magnetic fields retain a filamentary morphology, as initially volume-filling seed fields are compressed and amplified together with the collapsing large-scale structure \citep{zeldovich1970,kulsrud1997}. This behavior is not seen as clearly in the SN-only model, where magnetization is introduced locally by galactic winds and therefore does not naturally trace the same filamentary network\footnote{Note that although we use the terms `filaments' and 'filamentary network', a subset of our identified populations is thin and hosts no galaxies (as seen by the lack of white contours), and may not directly correspond to cosmic filaments that are typically discussed in the observational literature.}. 
In a realistic setup, we expect the Biermann \citep{biermann1950} and Durrive \citep{durrive2015} batteries to produce a feeble magnetization of these filamentary structures during the epoch of reionization, even in the absence of a cosmological seed field \citep[see also][]{garaldi2021}. 
The BH-driven models differ again: feedback-driven transport spreads magnetized material over larger regions around halos, reducing the visual contrast between the filamentary structures seen in the primordial case and the more locally injected magnetization characteristic of the SN-only run.

\begin{figure*}
    \centering
    \includegraphics[width=18cm]{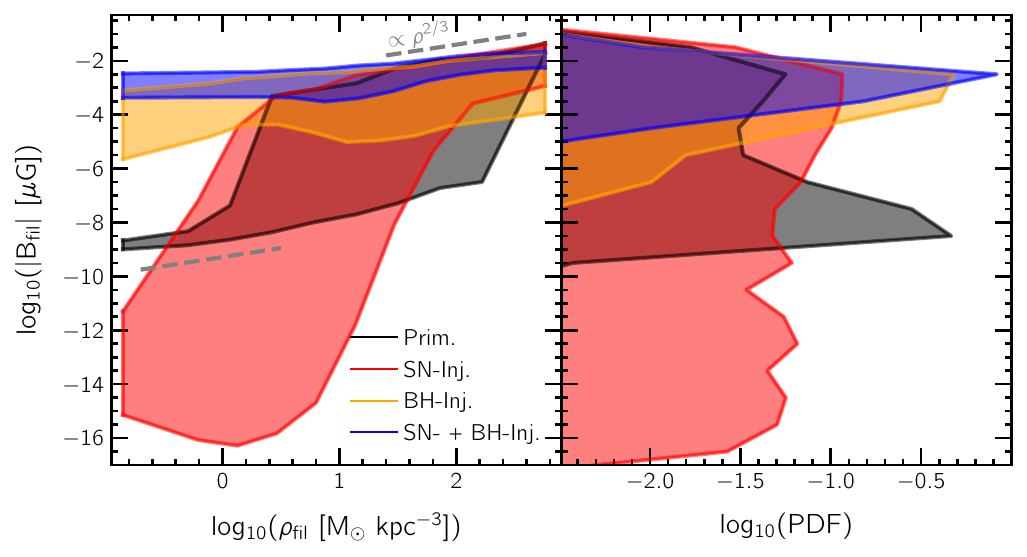}
    \caption{\textbf{Magnetic field strengths in filaments at $\mathbf{z=0}$:} Left: magnetic field strength as a function of density for the sample of filaments identified with \texttt{DisPerSE} {\protect \citep{sousbie2011}}, shown using 1$\sigma$-contours. The grey dashed line indicates the scaling corresponding to flux freezing, $|\rm{B}|$\,$\propto$\,$\rho^{2/3}$. Right: probability distribution functions of magnetic field strengths. In the primordial-only model, filaments separate into a low-field population that largely follows flux-freezing amplification of the primordial seed, and a high-field population where dynamo amplification has already boosted the field well above the primordial expectation; the intermediate branch marks systems in transition, where the field strengths are still growing. In the SN-only model, the $|\rm{B}|$\,$\propto$\,$\rho^{2/3}$ scaling is seen mainly in the densest filaments, while lower-density filaments show a comparatively broad distribution. In the BH-driven models, large-scale magnetic pollution is sufficiently widespread that filament magnetic field strengths turn more uniform across density.}
    \label{fig:filament_prop}
\end{figure*} 

To offer a more quantitative picture of the above trends, we begin by identifying cosmic filaments using \texttt{DisPerSE} \citep{sousbie2011}. Following a number of earlier works \citep[e.g.][]{cautun2013,ge2023,bahe2025}, we do so by (a) constructing the three-dimensional dark matter (DM) density field ($\rho_{\rm{DM}}$) on a Cartesian grid with 512 cells per dimension (corresponding to a cell size of $\sim$\,72\,kpc), using a standard SPH interpolation scheme with the kernel size set to the radius enclosing the nearest $64$\,$\pm$\,$4$ DM particles, (b) remapping the grid to one tracing the logarithm of the density, $\log_{10} \rho_{\rm{DM}}$, (c) smoothing the resulting field with a spherical Gaussian kernel of standard deviation set to three grid cells ($\sim$\,216\,kpc), and (d) constructing the Morse-Smale complex (MSC) of the smoothed density field. Filament spines are thereafter identified from the ascending manifold of the MSC, retaining only structures above a persistence threshold of $0.25$\footnote{Value set based on visual inspection.} and enforcing a loop-free topology. Finally, the resulting skeleton is smoothed over 10 iterations to suppress numerical fluctuations and produce a continuous representation of the cosmic web (see, e.g., \citealt{ge2024} for a comprehensive description of the various tunable parameters associated with \texttt{DisPerSE}).

In practice, the filaments as traced by \texttt{DisPerSE} are composed of a sequence of straight-line segments between a set of sampling points. We define the length of a filament (L$_{\rm{fil}}$) as the sum of the lengths of its segments, and in order to excise any remaining small-scale spurious features, discard those with L$_{\rm{fil}}$\,$<$\,$1000$\,kpc. To assign a characteristic density or magnetic field strength to each filament, we stack all gas cells within a cylinder of radius $500$\,kpc around each segment, with cells associated with any FoF halo discounted, and compute the median of this selection. Although not shown explicitly, we have experimented with apertures as small (large) as $200$ ($2000$) kpc, and mention that the results we present remain qualitatively unaffected.

In the left panel of Figure~\ref{fig:filament_prop}, we explore the relation between the density and magnetic field strengths of filaments, with shaded bands portraying the 16$^{\rm{th}}$-84$^{\rm{th}}$ percentile regions in the corresponding plane. The right panel plots the corresponding probability distribution function (PDF) of the field strengths. In both panels, colors correspond to the different seeding mechanisms, with the same scheme as earlier: black to the primordial-only run, red (orange) to only injection via SN- (BH-) winds, and blue to the run with both modes of winds active.

\begin{figure}
    \centering
    \includegraphics[width=9cm]{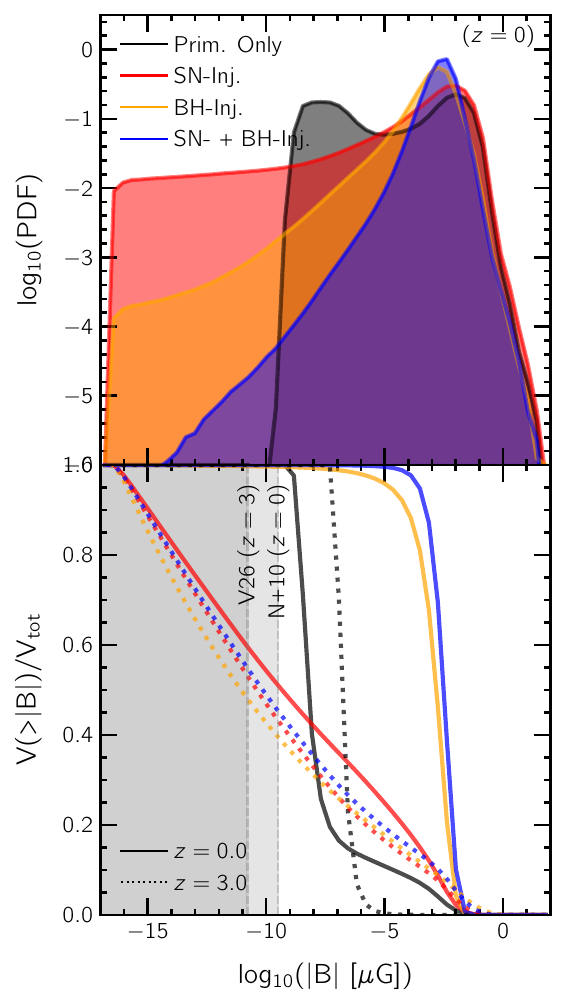}
    \caption{\textbf{Magnetic field distribution in the intergalactic medium:} Top: probability distribution functions of IGM magnetic field strengths at $z=0$. Bottom: cumulative volume filling fraction above a given magnetic field strength at $z=0$ (solid) and $z=3$ (dotted). Vertical reference lines, alongside shaded gray bands, indicate observational lower limits inferred from $\gamma$-ray cascade constraints (\protect{\citealt{neronov2010,vovk2026}}). SN-only injection underproduces widespread magnetization relative to the above-mentioned constraints, whereas BH-driven models yield stronger large-scale magnetic fields and larger filling fractions, broadly consistent with present-day constraints but remaining in mild tension at high redshifts.}
    \label{fig:igm_prop}
\end{figure}

Beginning with the primordial-only model, filaments largely separate into two distinct subsets. A low-field population, where the field strengths predominantly trace flux-freezing amplification of the initial primordial field ($|\rm{B}|$\,$\propto$\,$\rho^{2/3}$; dashed gray curve), constitutes the majority, as seen in the PDF in the right panel; these correspond to regions of zero/low stellar density, i.e volumes devoid of galaxies, also seen as thin features in Figure~\ref{fig:filament_vis} that have no white contours in their vicinity. On the other extreme, a high-field population exists, where amplification has already exponentially boosted the field beyond the initial value. An intermediate branch, comprising systems in transition where field strengths are presently growing, is comparatively rare. Overall, this relation is qualitatively similar to that obtained when considering all gas cells in the simulation volume \citep[e.g.][]{marinacci2018,garaldi2021}, with the low-, intermediate-, and high-field branches reflecting different stages of magnetic field amplification \citep[see also][]{dolag1999,dolag2002}. 

In the SN-only run, the above mentioned low-field population is essentially absent, and is instead replaced by filaments that are magnetized to varying degrees. As discussed in the context of Figure~\ref{fig:filament_vis}, this follows naturally from the absence of an initial primordial field: without a volume-filling seed field, magnetic field lines cannot be compressed together with the collapsing cosmic web, and filament magnetization must instead proceed through the transport of magnetic energy from nearby galaxies. In practice, SN-driven winds are relatively inefficient at polluting the most diffuse environments, resulting in a broad population of weakly magnetized filaments at low densities. At higher densities, however, a distinct high-field population emerges whose magnetic field strengths approach those seen in the primordial-only run and follow a similar scaling with density. These systems likely correspond to filaments that have experienced more efficient enrichment by galactic outflows, allowing their magnetic field strengths to reach levels comparable to those attained in the presence of an initial primordial seed field.

The behavior changes qualitatively once BH-driven winds are included. In both the BH-only and combined SN+BH runs, the magnetic field strengths exhibit only a weak dependence on filament density, with the median relation largely independent of $\rho_{\rm{fil}}$, i.e. nearly flat across much of the dynamic range shown. This is a direct consequence of the large spatial extent of BH-driven outflows, which distribute magnetic energy far beyond the immediate vicinity of galaxies. Consistent with the visual impression from Figure~\ref{fig:filament_vis}, the resulting magnetization is comparatively homogeneous, reducing the distinction between low- and high-density filaments and producing a substantially narrower range of field strengths at fixed density.

\subsection{Properties of IGM gas}\label{ssec:igm_gas}

While the preceding analysis focused on magnetic fields within individual filaments, it is also useful to consider the global magnetization of the diffuse IGM. In Figure~\ref{fig:igm_prop}, we therefore examine the distribution of magnetic field strengths of all gas outside collapsed structures. The top panel presents the corresponding PDFs at $z$\,$=$\,$0$, while the bottom panel shows the cumulative volume filling fraction above a given field strength at $z$\,$=$\,$0$ (solid) and $z$\,$=$\,$3$ (dotted). As before, colors denote the adopted seeding prescription(s).

Several trends are apparent. In the primordial-only run, the IGM field distribution is relatively narrow at the weakest field strengths, reflecting the presence of an initially volume-filling seed field that is subsequently compressed and amplified by structure formation. The SN-only model behaves quite differently: while a significant fraction of the IGM remains extremely weakly magnetized, producing an extended low-field tail in the PDF, the distribution also reaches field strengths comparable to the other models in regions that have been enriched by galactic outflows. This broad distribution reflects the highly inhomogeneous nature of SN-driven magnetization, which can strongly magnetize gas near galaxies and dense structures but does not efficiently fill the most diffuse regions of the IGM. In contrast, the BH-driven models produce much larger volume filling fractions at intermediate-to-high field strengths, indicating that BH-driven outflows distribute magnetic energy over substantially larger regions of the intergalactic medium (see also Section~\ref{ssec:disc} and Appendix~\ref{sec:app_bh_wind}).

The above differences are even more evident in the cumulative volume filling fractions shown in the bottom panel. At $z$\,$=$\,$0$, the primordial-only run predicts that roughly half of the IGM volume is magnetized above $\sim$\,$10^{-8}$\,$\mu$G, with the filling fraction declining rapidly towards stronger fields. The SN-only model exhibits a much more gradual decline, consistent with the broad PDF discussed above, but fails to magnetize a sufficiently large fraction of the volume to field strengths inferred from $z=0$ $\gamma$-ray cascade constraints (\citealt{neronov2010}; dashed vertical line at $\sim$\,$10^{-9.5}$\,$\mu$G). In contrast, both BH-driven models maintain filling fractions close to unity up to substantially larger field strengths, indicating that magnetic energy is distributed over a much larger fraction of the intergalactic volume\footnote{Although not explicitly shown here, we mention that the total magnetic energy contained within IGM gas is also higher in the runs with BH-winds active (at $z$\,$=$\,$0$). The run with SN-winds only is relatively the least, and the primordial-only case is intermediate in this regard.}. The combined SN+BH run is particularly efficient in this regard, yielding the highest filling fractions across nearly the entire dynamic range explored.

The redshift evolution of the filling fractions further highlights the different timescales associated with each magnetization channel. In the astrophysical-only models, the volume occupied by strongly magnetized gas decreases substantially towards earlier epochs, reflecting the finite time required for magnetic fields to be generated in galaxies and transported into low-density environments by feedback-driven outflows. This effect is most pronounced for the BH-only model, but remains visible even in the SN-driven runs. Consequently, although the BH-driven models broadly satisfy present-day lower limits inferred from $\gamma$-ray cascade measurements, they remain in mild tension with the corresponding constraints at $z\sim3$ (\citealt{vovk2026}; dashed vertical line at $\sim$\,$10^{-10.8}$\,$\mu$G). By contrast, the primordial-only run evolves more weakly in volume filling fraction, owing to the presence of an initially volume-filling seed field that is already embedded in the diffuse IGM prior to structure formation.

Although we compare our results with the above-discussed $\gamma$-ray cascade constraints, we here offer a word of caution: they primarily attribute the energy loss of the particle pairs to inverse Compton scattering, while plasma instabilities may contribute, or even dominate, the energy dissipation budget \citep[see e.g.][]{broderick2012,perry2021}, making it non-trivial to assess their robustness. In either case, the magnetic field distribution in the diffuse IGM remains strongly dependent on the adopted seeding prescription. Differences that are largely erased within collapsed halos persist in low-density environments, leading to substantially different field-strength distributions and volume filling fractions at both low and high redshift.

\subsection{Magnetic coherence lengths}\label{ssec:coh_len}

As a final analysis, we examine the coherence scales of magnetic field lines across the different magnetogenesis models. Previous studies have often characterized coherence through correlation or integral scales derived from the magnetic power spectrum \citep[e.g.][]{enslin2003,subramanian2006}. While such approaches provide a useful global measure of magnetic structure, they do not readily distinguish between different environments.

Here, we instead adopt a local, field-line based diagnostic. Using the \texttt{THOR} code \citep{byrohl2025}, we launch rays from a set of $10^6$ randomly selected points\footnote{Note that this corresponds to $\sim$\,$1$\,$\%$ of the gas cells in each run. This large but sparse sampling provides good coverage of the simulation volume while reducing the likelihood of repeatedly tracing the same local field-line segments.} throughout the simulation domain and propagate them along the local magnetic field direction\footnote{For a visualization movie on how these rays propagate, see the \texttt{THOR} documentation \href{https://thor-rt.org/cookbook/analyses/coherence-length/}{website}.}. We then define a proxy for the coherence length, L$_{\rm c,\vec{B},30^\circ}$, as the distance travelled before the field orientation between successive cells deviates by more than $30^\circ$. Measurements involving fewer than 10 cell crossings are discarded to reduce numerical noise. Although the above choices of $30^\circ$ and 10 cells are rather arbitrary, we have verified that varying these thresholds do not strongly alter the trends in any qualitative manner.

\begin{figure}
    \centering
    \includegraphics[width=9cm]{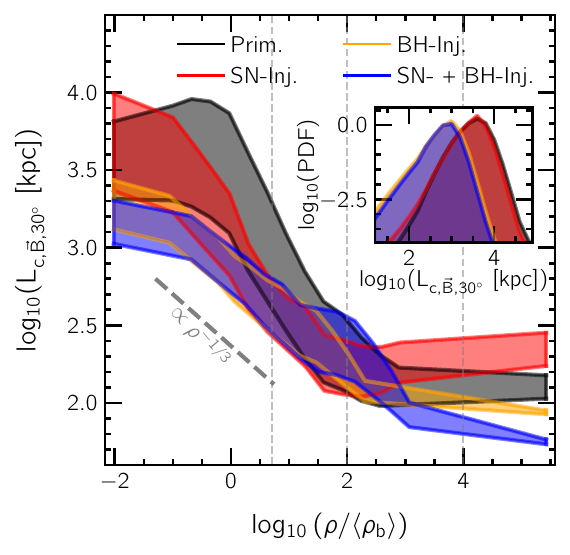}
    \caption{\textbf{Assessing the coherence of magnetic field lines:} Proxy for the magnetic coherence length, L$_{\rm c,\vec{B},30^\circ}$, as a function of gas overdensity at $z$\,$=$\,$0$. Shaded regions indicate the 16$^{\rm th}$-84$^{\rm th}$ percentile range. Colors correspond to the various seeding models, following the same scheme as earlier figures. The dashed gray line shows a scaling of $\propto$\,$\rho^{-1/3}$, corresponding approximately to the density dependence of the spatial resolution. The inset shows the PDFs of L$_{\rm c,\vec{B},30^\circ}$ measured across all sampled locations for the different runs. At fixed overdensity, differences between the various magnetogenesis models are apparent, as discussed further in Section~\ref{ssec:coh_len}.}
    \label{fig:coherence_length}
\end{figure}

Figure~\ref{fig:coherence_length} summarizes the resulting coherence length distributions. The main panel shows L$_{\rm c,\vec{B},30^\circ}$ as a function of gas overdensity, $\rho/\langle\rho_b\rangle$\footnote{We define $\rho$ as the median density of the selection of gas cells that the ray propagates through.}, at $z$\,$=$\,$0$, while the inset plots the corresponding PDFs obtained from all sampled locations. For reference, the vertical dashed lines at $\log_{10}(\rho/\langle\rho_b\rangle)$\,$\sim$\,0.7, 2, and 4 roughly separate gas residing in the most under-dense regions outside halos, in cosmic filaments and sheets, within halos, and in the central regions of halos, respectively \citep[e.g.][]{dolag2005,marinacci2018}. The scaling of $\propto$\,$\rho^{-1/3}$, shown by the dashed gray line, is approximately the density dependence of spatial resolution. Consequently, care should be taken when comparing coherence lengths across large density ranges. Instead, the more meaningful comparisons are between different magnetogenesis models at fixed overdensity.

In the most diffuse gas ($\rho/\langle\rho_b\rangle$\,$\lesssim$\,$1$), the primordial-only and SN-only models extend to the largest coherence lengths, although the percentile ranges overlap across all runs. At intermediate overdensities ($1$\,$\sim$\,$\rho/\langle\rho_b\rangle$\,$\lesssim$\,$100$), the primordial-only run remains shifted towards larger L$_{\rm c,\vec{B},30^\circ}$ than the astrophysical-only variations, consistent with the presence of initially volume-filling field lines that are subsequently compressed during structure formation. The ordering changes in the densest gas ($\rho/\langle\rho_b\rangle$\,$\gtrsim$\,$100$): the SN-only model exhibits the largest coherence lengths, while both BH-driven models have systematically smaller values. This suggests that BH-driven transport not only spreads magnetic fields over larger volumes, but also produces a more rapidly varying field-line geometry in all environments, likely reflecting magnetization by repeated, localized outflow events with varying field orientations.

The inset provides a complementary view by integrating over all sampled environments. Interestingly, the primordial-only and SN-only models yield remarkably similar coherence-length distributions, despite their very different magnetic field strength distributions (Figure~\ref{fig:igm_prop}), and the lack of a uniform seed field in the latter. This similarity should be however interpreted with some caution: it appears to arise from a compensation across the overdensity relation, with the primordial-only run reaching larger coherence lengths in diffuse and intermediate-density gas, and the SN-only run doing so in the densest regions where the fields are injected and shaped by the local galactic flow. By contrast, both BH-driven models are shifted towards systematically smaller coherence lengths by $\sim$\,$0.5$\,dex, consistent with the noticeable vertical offset in the main panel.

Importantly, such differences are not only of theoretical interest, but also have observational consequences: many proposed probes of intergalactic magnetic fields, including ultra-high-energy cosmic rays and $\gamma$-ray induced pair cascades, depend not only on the field strength but also on the characteristic scale over which magnetic field directions remain correlated \citep[e.g.][]{neronov2010,durrer2013}. Measurements sensitive to both quantities may therefore provide additional leverage for distinguishing between competing magnetogenesis scenarios. In this regard, the coherence scale of magnetic fields may preserve information about the origin and transport history of cosmic magnetism that is not readily apparent from magnetic field strengths alone.

\subsection{Outlook}\label{ssec:disc}
A long-standing question in studies of cosmic magnetism is whether the observed magnetic field distribution of the Universe can be explained entirely through astrophysical processes, or whether an additional primordial contribution is required \citep[e.g.][]{widrow2002,durrer2013,subramanian2016,garaldi2021}. Over the past decade, increasingly sophisticated numerical studies have demonstrated that astrophysical processes, including galactic winds and SMBH-driven outflows, are capable of both injecting magnetic fields into the surrounding gas and redistributing them far beyond their host galaxies \citep[e.g.][]{bertone2006,donnert2009,beck2013,marinacci2015}. At the same time, the predicted level of magnetization remains strongly model dependent. %The results presented in this work broadly reinforce this picture, demonstrating that different astrophysical magnetogenesis channels can produce substantially different magnetic field distributions despite operating within the same underlying galaxy formation model.

In this regard, it is instructive to place our findings in context of \citet{vazza2017}, who likewise explored a suite of primordial and astrophysical magnetogenesis models in cosmological simulations, albeit with different numerical prescriptions for both the seeding and transport of fields. While their study and ours arrive at largely similar conclusions regarding the viability of astrophysical magnetization, \citet{vazza2017} predict a more modest enhancement of intergalactic magnetic fields than found here, especially in the case of SMBH-driven magnetization. Although a direct comparison is complicated by differences in numerical resolution and subgrid modelling, the contrast highlights the extent to which predictions for astrophysical magnetogenesis remain coupled to other details of galaxy formation physics. 

Beyond the above qualitative comparison, the model variations explored in Appendix~\ref{sec:app_bh_wind} provide additional quantitative insight into the physical processes that regulate astrophysical magnetogenesis. An important takeaway is that the efficiency of magnetization is not determined solely by the amount of magnetic energy explicitly injected into the gas. For example, varying the magnetic energy fraction associated with SMBH feedback produces only comparatively modest changes in the resulting intergalactic magnetic field distribution (Appendix~\ref{ssec:kin_vs_mag_inj}). By contrast, modifications that affect the propagation of feedback-driven outflows, including changes to the burstiness of feedback events (Appendix~\ref{ssec:bh_wind_burst}), wind velocities (Appendix~\ref{app:wind_vel}), and the timing of SMBH growth (Appendix~\ref{ssec:app_bh_seed}), can have a substantially larger impact. This suggests that, once a magnetized reservoir has been established within galaxies, the subsequent redistribution of magnetic fields into diffuse environments becomes at least as important as the direct injection process itself. In this picture, our specific BH-wind driven outflows act not only as sources of magnetic energy, but also as a mechanism for coupling magnetized galactic gas to the wider cosmic web.

Within the framework explored here, the most significant challenge for purely astrophysical magnetogenesis models is not reproducing the present-day magnetic field distribution, but rather achieving sufficient magnetization of the intergalactic medium at earlier cosmic epochs. The parameter variation presented in Appendix~\ref{ssec:app_bh_seed} suggests that earlier SMBH growth can potentially increase magnetic field filling fractions at high redshift. Although this explored run produces only a modest change, it suggests that the onset and subsequent evolution of SMBH feedback constitute important degrees of freedom that are yet to be fully explored. This is particularly relevant in light of recent simulation efforts that seed SMBHs at earlier epochs and in lower-mass halos than typically assumed \citep[e.g.][]{bowmick2024}, potentially extending the period over which SMBH-driven magnetization can operate. 

More broadly, future efforts may benefit from treating intergalactic magnetic fields as an additional calibration target alongside more traditional observables such as the stellar mass function, cosmic star formation history, and black hole scaling relations \citep{pillepich2018}. While the present work focuses on maximizing the impact of astrophysical magnetogenesis, a more ambitious goal would be to identify feedback models that simultaneously reproduce both galaxy formation observables and the evolving magnetic field distribution discussed above, along with related measures of IGM magnetization such as the volume filling fraction and line-of-sight covering fraction of magnetized outflow bubbles, which directly probe the efficiency and extent of feedback-driven transport of magnetic fields into the cosmic web \citep[e.g.][]{ag2021}. Whether such a calibration can be achieved within a purely astrophysical framework, or ultimately requires an additional primordial contribution, remains an open question.

\section{Summary and conclusions}\label{sec:summary}

We present a suite of cosmological magnetohydrodynamical simulations (L$_\mathrm{box}$\,$=$\,$25$ Mpc/h) run with the code \texttt{AREPO}, based on the IllustrisTNG model, augmented with additional prescriptions for magnetic field injection associated with stellar feedback and SMBH activity. By comparing primordial, SN-driven, BH-driven, and hybrid seeding scenarios, we investigate how different magnetogenesis channels shape the magnetic properties of halos, the circumgalactic medium, and the intergalactic medium. Our main findings are as follows:

\begin{enumerate}

\item Halo magnetic field strengths at $z$\,$=$\,$0$ are largely insensitive to the origin of the seed field. Across halos with masses M$_{200c}$\,$\sim$\,$10^{11.5}-10^{12}$\,M$_{\odot}$, radial magnetic field profiles converge to similar amplitudes irrespective of whether the initial magnetization is primordial or astrophysical in origin. Other halo gas properties, particularly metallicity and thermodynamic structure, retain a clearer imprint of the underlying feedback model (Figure~\ref{fig:radial_prof}).

\item Astrophysical magnetic injection primarily affects the onset of magnetic amplification rather than the final saturated state within haloes. Magnetic power spectra show that feedback-driven injection, especially through SN-driven winds, accelerates the growth of magnetic energy on resolved scales and promotes earlier saturation of the small-scale dynamo. By $z$\,$=$\,$0$, however, the magnetic energy budget within halos is regulated predominantly by dynamo amplification, erasing much of the memory of the original seeding mechanism (Figure~\ref{fig:power_spectra}).

\item The diffuse gas outside halos, including gas populating filamentary bridges between halos, retains a substantially stronger memory of magnetogenesis than the halo population itself. As opposed to magnetic field strengths within halos converging across seeding prescriptions, the spatial distribution and volume-filling fraction of magnetic fields in the intergalactic medium remains strongly dependent on the adopted injection model (Figures~\ref{fig:filament_prop} and \ref{fig:igm_prop}).

\item In our numerical scheme, SN-driven injection alone is insufficient to reproduce inferred lower limits of intergalactic magnetic field strengths. While stellar feedback magnetizes the surrounding environment efficiently, it fails to generate field strengths across the volume filling IGM to levels consistent with constraints from $\gamma$-ray cascade measurements, both at $z$\,$=$\,$0$ and at higher redshifts. On the other hand, our BH-wind scheme is considerably more effective at magnetizing low-density gas and can satisfy present-day observational constraints on intergalactic magnetic fields. Nevertheless, mild tension remains with inferred field strengths at $z$\,$\sim$\,$3$, suggesting that in order to reconcile with these specific measurements, either an additional primordial source of magnetization may still be required, or a modified feedback prescription (Figure~\ref{fig:igm_prop}).

\item Magnetic field strengths and magnetic field topology do not necessarily evolve in tandem. While BH-driven models produce the strongest and most volume-filling intergalactic magnetic fields, they also exhibit systematically smaller magnetic coherence lengths than either primordial-only or SN-driven scenarios. Conversely, the primordial-only and SN-driven models yield remarkably similar coherence-length distributions despite their differing magnetic field strengths. These results suggest that magnetic field topology may preserve additional information about the origin and transport history of cosmic magnetism beyond that encoded in field strengths alone (Figure~\ref{fig:coherence_length}).

\end{enumerate}

All in all, our results suggest that the different astrophysical magnetization scenarios considered here, including the associated large-scale transport by kinetic energy injections, can play a major role in establishing magnetized reservoirs within galaxies and their surrounding environments. However, whether they can alone account for the magnetic field distribution of the entire volume-filling IGM, particularly at high-$z$, remains an open question. Future work will explore a broader range of feedback prescriptions and SMBH models, with the aim of simultaneously reproducing intergalactic magnetic field constraints across cosmic time and other observational benchmarks of galaxy formation.

\section*{Data Availability}
Data directly related to this publication is available upon reasonable request to the corresponding author. This work has benefitted from the \texttt{scida} library\footnote{\url{https://github.com/cbyrohl/scida}} \citep{byrohl2024} for data analysis, and the \texttt{temet} package\footnote{\url{https://github.com/dnelson/temet}} \citep{nelson2025} for the various visualization figures presented.

\begin{acknowledgements}
RR and EG acknowledge support by the World Premier International Research Center Initiative (WPI), MEXT, Japan. We thank Annalisa Pillepich for sharing relevant data points, Rüdiger Pakmor, Volker Springel and Dylan Nelson for insightful feedback, Daniela Galárraga-Espinosa for useful suggestions on using \texttt{DisPerSE}, and Ivegen Vovk for discussions on intergalactic fields. This work has made use of the \texttt{idark} and \texttt{gw} clusters at IPMU, the \texttt{xd2000} system operated by the NAOJ, and of NASA's Astrophysics Data System Bibliographic Services.
\end{acknowledgements}

\bibliographystyle{mnras}
\bibliography{references}

\newpage

\appendix

\section{Resolution convergence}\label{sec:app_res_conv}

\begin{figure*}
    \centering
    \includegraphics[width=18cm]{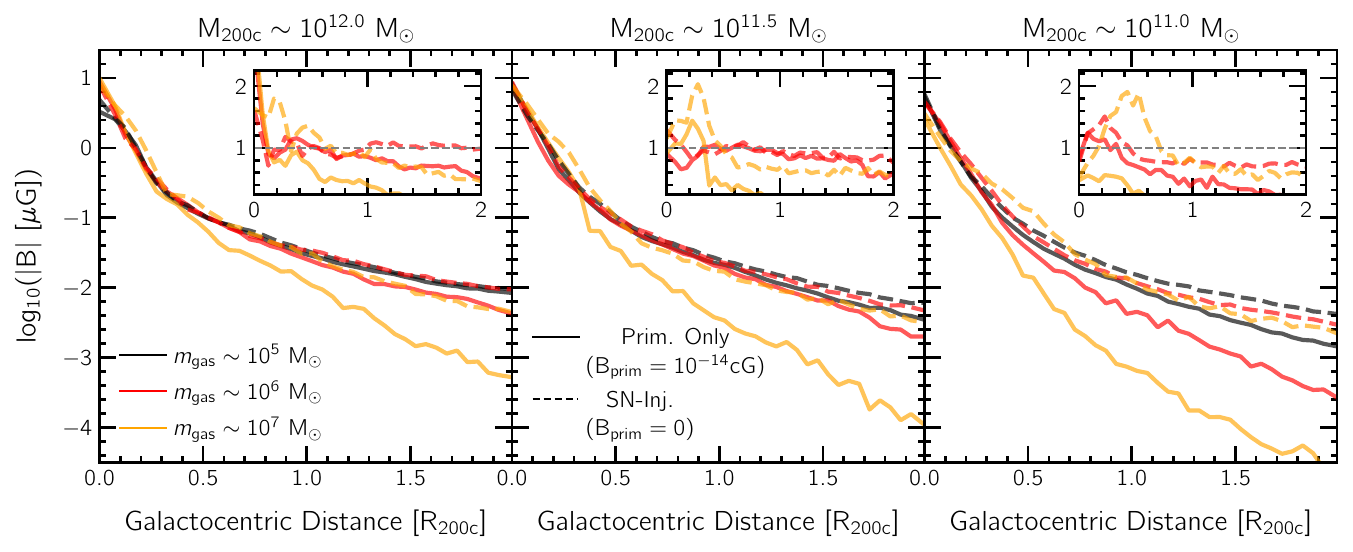}
    \caption{\textbf{Assessing numerical resolution convergence of halo magnetic field profiles:} Spherically-averaged radial magnetic field profiles are shown for halos in a 12.5\,Mpc/h box with baryonic mass resolutions of $m_{\rm{gas}}$\,$\sim$\,$10^5$\,M$_\odot$ (black), $10^6$\,M$_\odot$ (red), and $10^7$\,M$_\odot$ (orange). Solid lines correspond to the primordial-only model, while dotted lines show the SN-only injection model. From left to right, columns present halos with M$_{\rm{200c}}$\,$\sim$\,$10^{12}$, $10^{11.5}$, and $10^{11}$\,M$_\odot$, respectively. Insets show the ratio of each profile relative to the highest-resolution run. The SN-only model approaches numerical convergence more rapidly than the primordial-only case, consistent with feedback-driven injection accelerating the onset of small-scale dynamo amplification.}
    \label{fig:app_res_conv_rad_prof}
\end{figure*}

\begin{figure}
    \centering
    \includegraphics[width=8cm]{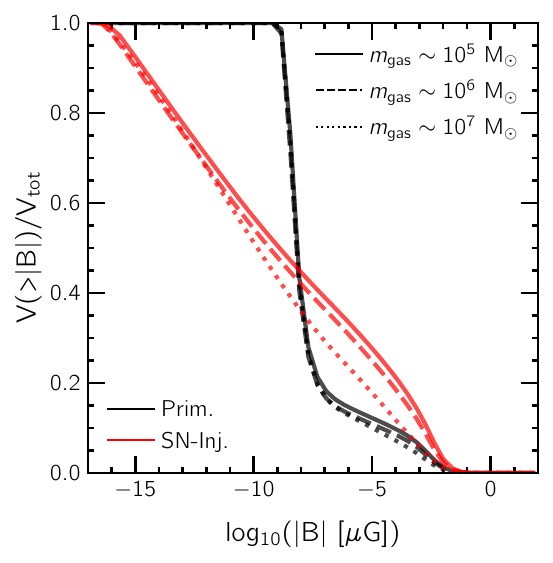}
    \caption{\textbf{Resolution dependence of intergalactic magnetic field filling fractions:} Cumulative volume filling fractions above a given magnetic field strength are shown for simulations with baryonic mass resolutions of $m_{\rm{gas}}$\,$\sim$\,$10^5$, $10^6$, and $10^7$\,M$_\odot$, indicated by different line styles, from the same 12.5\,Mpc/h box as Figure~\ref{fig:app_res_conv_rad_prof}. Colors distinguish the primordial-only and SN-only injection suites. While resolution dependence persists at the weakest field strengths, the fiducial resolution adopted throughout the main text ($m_{\rm{gas}}$\,$\sim$\,$10^6$\,M$_\odot$) yields largely converged IGM magnetic field filling fractions.}
    \label{fig:app_res_conv_igm_b}
\end{figure}

The amplification and transport of magnetic fields in cosmological simulations are known to depend on numerical resolution, since both the efficiency of small-scale dynamo amplification and the ability to resolve magnetized outflows improve as increasingly smaller scales become accessible \citep{pakmor2024}. Throughout the main text, results presented are derived from simulations adopting a fiducial baryonic mass resolution of $m_{\rm gas}$\,$\sim$\,$10^6$\,M${_\odot}$. In this appendix, we quantify the residual resolution dependence of our results using a suite performed in a smaller 12.5\,Mpc/h box at three different mass resolutions, $m_{\rm gas}$\,$\sim$\,$10^5$, $10^6$, and $10^7$\,M${_\odot}$.

Figure~\ref{fig:app_res_conv_rad_prof} examines the convergence of spherically averaged magnetic field profiles within halos at $z$\,$=$\,$0$. We compare the primordial-only model (solids lines) to the SN-driven injection model (dashed curves) across three halo mass bins spanning M$_{200{\rm c}}$\,$\sim$\,$10^{11}$-$10^{12}$\,M${_\odot}$, shown in descending order from left to right. As expected, convergence improves with increasing halo mass, reflecting the larger number of better-resolved gas cells available to capture turbulent amplification and feedback-driven transport processes. For halos of M$_{200{\rm c}}$\,$\gtrsim$\,$10^{11.5}$\,M${_\odot}$, both seeding models exhibit reasonably converged radial profiles at our fiducial resolution, with differences relative to the highest-resolution runs typically limited to factors close to unity across most radii (inset). At lower halo masses, however, the primordial-only model displays substantially stronger resolution dependence, particularly in the outskirts of halos where magnetic fields are weakest.

A notable result is that the SN-driven injection model approaches convergence significantly more rapidly than the primordial-only case. This behaviour is consistent with the trends identified in Figure~\ref{fig:power_spectra} of the main text, where feedback-driven injection accelerates the onset of magnetic field growth on resolved scales. In the primordial-only scenario, magnetization of halo gas relies on the transport and subsequent amplification of initially weak magnetic fields, making the results particularly sensitive to the resolution-dependent efficiency of the small-scale dynamo. By contrast, direct magnetic injection through galactic winds seeds magnetic energy on scales that are already resolved, and with a larger seed field, reducing the dependence on numerical amplification and yielding more stable halo magnetic field profiles across resolutions.

We next consider the impact of resolution on intergalactic magnetic fields. Figure~\ref{fig:app_res_conv_igm_b} shows cumulative volume filling fractions above a given magnetic field strength for both the primordial-only (black) and SN-injection (red) suites. The strongest magnetic fields exhibit relatively weak resolution dependence, while differences become increasingly apparent around relativelty weaker fields ($\sim$\,$10^{-6}$\,$-$\,$10^{-5}$\,$\mu$G). This behaviour is expected because these weaker fields predominantly occupy diffuse intergalactic environments that are most sensitive to the details of numerical mixing and transport (Section~\ref{ssec:igm_gas}).

Importantly, the fiducial resolution adopted throughout this work ($m_{\rm gas}$\,$\sim$\,$10^6$\,M${_\odot}$) produces filling fractions that lie close to the corresponding high-resolution results over the majority of the magnetic field distribution. Although complete convergence is not achieved at the extreme low-field tail, the remaining differences are modest compared to the order-of-magnitude variations introduced by different seeding prescriptions. We therefore conclude that the principal results presented in the main text are robust with respect to numerical resolution, while noting that magnetic field strengths in halos below $M_{200{\rm c}}\sim10^{11.5}{\rm M_\odot}$ remain subject to appreciable convergence uncertainties, particularly in the primordial-only model.

\section{On the BH-Wind scheme}\label{sec:app_bh_wind}

\begin{figure*}
    \centering
    \includegraphics[width=18cm]{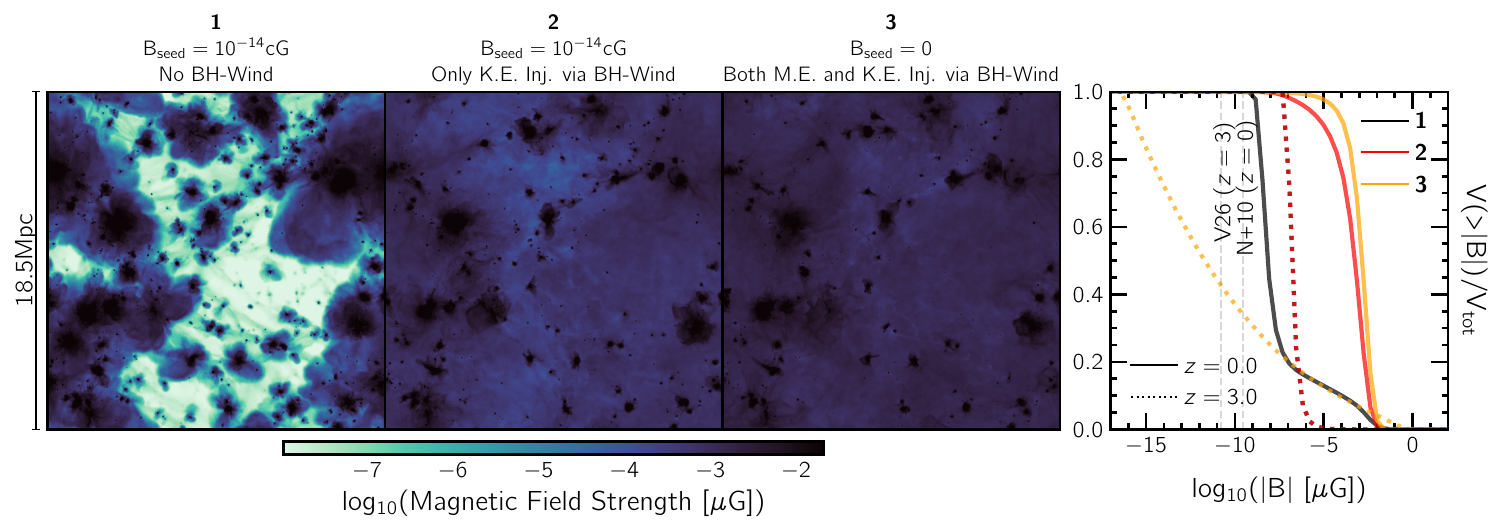}
    \caption{\textbf{Role of kinetic and magnetic energy injection in BH-driven wind models:} Left: projected magnetic field maps comparing three scenarios: primordial-only seeding, BH winds carrying kinetic energy only, and BH winds carrying both kinetic and magnetic energy. Right: corresponding cumulative volume filling fractions of magnetic field strengths at $z$\,$=$\,$0$ and $z$\,$=$\,$3$. The large-scale magnetization of the IGM is driven primarily by kinetic energy injection from BH winds, which transports magnetized material into diffuse environments; explicit magnetic energy loading provides an additional, more moderate boost to the resulting field strengths and filling fractions.}
    \label{fig:app_bh_wind}
\end{figure*}

\begin{figure*}
    \centering
    \includegraphics[width=18cm]{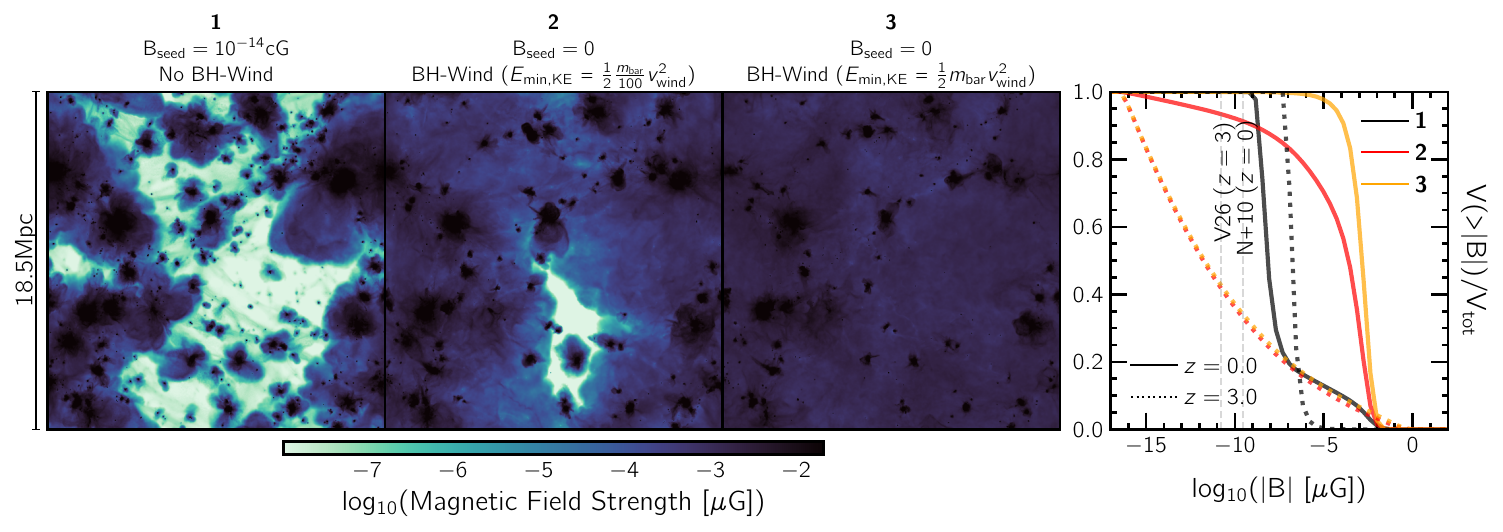}
    \caption{\textbf{Effect of burstiness of the injection:} Following the layout of Figure~\ref{fig:app_bh_wind}, we compare BH-driven wind models with different minimum kinetic energy thresholds per wind event. Less bursty feedback results in less efficient large-scale transport of magnetized material and consequently lower magnetic field filling fractions than in the fiducial model.}
    \label{fig:app_bh_wind_burst}
\end{figure*}

\begin{figure*}
    \centering
    \includegraphics[width=18cm]{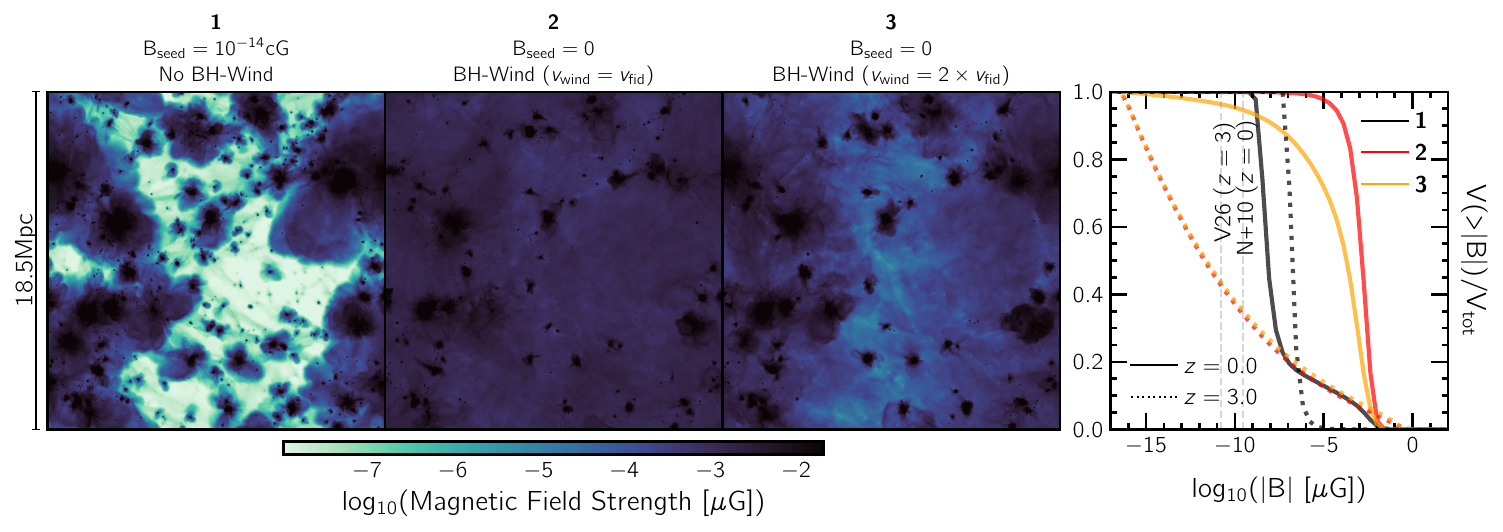}
    \caption{\textbf{Impact of varying the launch velocity of the wind:} We compare the fiducial BH-driven wind model to a variation in which the wind launch velocity is increased by a factor of two. Contrary to the naive expectation that faster winds should produce more widespread magnetization, the higher-velocity model yields systematically lower magnetic field filling fractions, as described further in the text (Appendix~\ref{app:wind_vel}).}
    \label{fig:app_bh_wind_vel}
\end{figure*}

\begin{figure*}
    \centering
    \includegraphics[width=18cm]{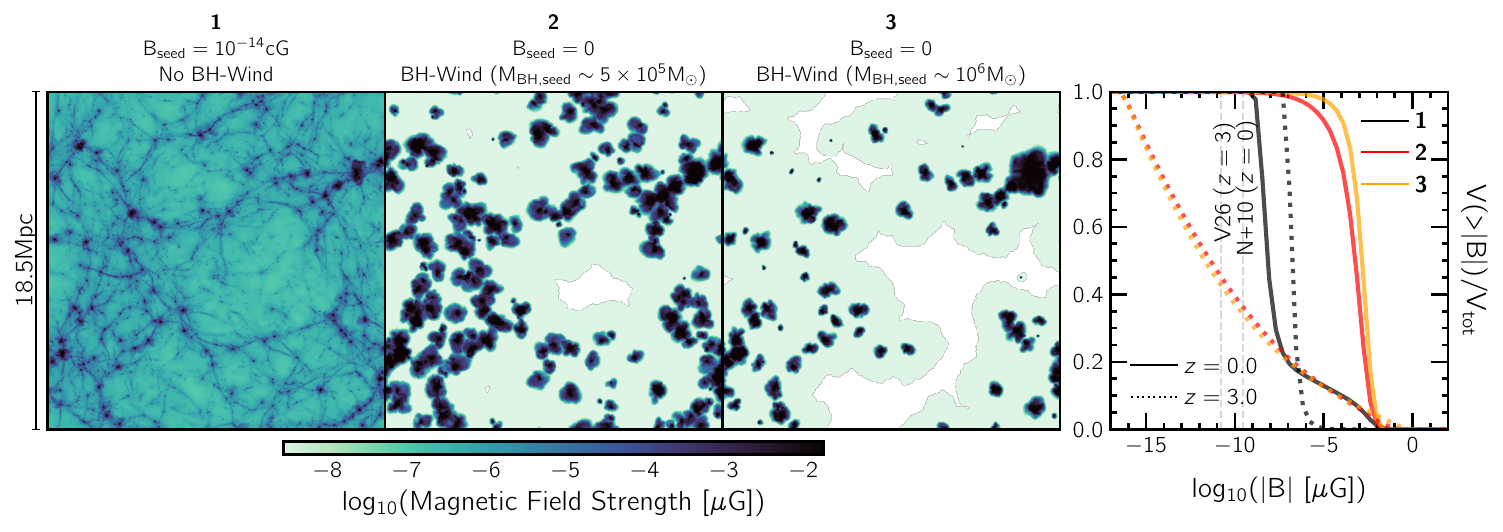}
    \caption{\textbf{Seeding smaller SMBHs at earlier cosmic epochs:} With a  similar layout as Figure~\ref{fig:app_bh_wind}, but showing projected magnetic field maps at $z$\,$=$\,$3$ instead of $z$\,$=$\,$0$ (and with a different colorbar), we compare BH-driven wind models with different SMBH seed masses. Earlier seeding of lower-mass SMBHs only leads to an increase, albeit rather marginal, in the magnetic field filling fraction at high redshift by allowing magnetized outflows to begin operating sooner.}
    \label{fig:app_bh_seed_mass}
\end{figure*}

The results presented in the main text suggest that SMBH-driven winds are substantially more effective than SN-driven winds at magnetizing the diffuse intergalactic medium. However, the efficiency of this process is expected to depend on the details of the underlying feedback model. In particular, the extent to which magnetic fields are transported into low-density environments may be sensitive to the burstiness, velocity, and onset of the outflows responsible for carrying magnetized gas away from galaxies.

In this appendix, we therefore explore a small set of variations, with a suite of 12.5\,Mpc/h at the same fiducial resolution of $m_{\rm gas}$\,$\sim$\,$10^6$\,M${_\odot}$, to the fiducial SMBH-driven transport prescription. Our goal is not to perform a systematic parameter study, but rather to identify which aspects of the feedback model have the largest impact on large-scale magnetization. In all but Appendix~\ref{ssec:kin_vs_mag_inj}, we keep the magnetic and kinetic energy fractions associated with the SMBH-driven winds fixed at their fiducial values ($\tau_{\rm mag,BH}$ and $\tau_{\rm kin,BH}$; Section~\ref{ssec:wind_calib}) and vary only the properties that govern how the injected energy is distributed in space and time. We focus on the same diagnostic introduced in Section~\ref{ssec:halo_gas}: the distribution of magnetic field strengths in the diffuse intergalactic medium.

Unless stated otherwise, all figures in this appendix follow a common layout. The left-most panel shows the primordial-only reference model, while the remaining panels show different SMBH-driven injection variants. The projected magnetic field maps provide a qualitative illustration of the resulting large-scale magnetization pattern, while the accompanying cumulative volume filling fractions quantify the extent to which magnetic fields permeate the intergalactic medium.

\subsection{Role of kinetic versus magnetic energy injection}\label{ssec:kin_vs_mag_inj}
In our fiducial BH-wind model, a fraction of the SMBH feedback energy is deposited directly in magnetic form, while the remainder is injected kinetically. An important question is therefore whether the enhanced magnetization of the diffuse intergalactic medium arises primarily from the explicit magnetic energy injection itself, or from the ability of the associated outflows to transport magnetized material away from galaxies. To assess the relative importance of these two channels, Figure~\ref{fig:app_bh_wind} compares variations with different values of $\tau_{\rm mag,BH}$ while leaving the remainder of the feedback model unchanged.

Several trends are apparent. First, the projected magnetic field maps show that varying $\tau_{\rm mag,BH}$ does modify the characteristic field strengths in the immediate vicinity of galaxies. However, the resulting differences become somewhat weaker when considering the cumulative volume filling fractions. In particular, the large-scale distribution of magnetic fields remains broadly similar between the two runs with BH-winds active, despite the lack of magnetic energy in one of the variations.

This behaviour suggests that the magnetization produced by the BH-driven channel is not set solely by the explicit magnetic energy budget of the wind. Instead, the large-scale field distribution appears comparatively insensitive to the precise value of $\tau_{\rm mag,BH}$, provided that some level of magnetic injection is present. The primary role of the SMBH-driven outflow in inducing a large-scale pollution of fields is therefore the redistribution of magnetized material into diffuse environments, rather than simply acting as a source of magnetic energy.

\subsection{Burstiness of the injection}\label{ssec:bh_wind_burst}

Our implementation of the BH-wind model accumulates feedback energy until a minimum threshold is reached, after which a wind event is launched. As a result, energy injection proceeds in a bursty manner, with relatively infrequent but energetic outflows. To explore the importance of this assumption, Figure~\ref{fig:app_bh_wind_burst} compares the fiducial model to a variation in which the minimum energy threshold is reduced by a factor of $100$, producing a more continuous sequence of wind events.

Despite the total injected energy remaining largely unchanged (not shown explicitly), the resulting magnetization differs significantly. The less bursty model produces systematically lower magnetic field filling fractions and weaker large-scale magnetization patterns than the fiducial run. This suggests that a small number of energetic events is more effective at transporting magnetic fields into diffuse environments than a larger number of weaker outflows.

The dependence on burstiness indicates that the temporal structure of the feedback plays an important role in regulating the efficiency of large-scale magnetization. In particular, the ability of individual wind events to transport magentized gas far from their host galaxies appears to be at least as important as the total amount of energy injected over cosmic time.

\subsection{Varying the launch velocity}\label{app:wind_vel}

A natural expectation is that faster outflows should transport magnetized material to larger distances and therefore produce more widespread magnetization of the diffuse intergalactic medium. To test this, Figure~\ref{fig:app_bh_wind_vel} compares the fiducial BH-driven model to a variation in which the wind launch velocity is increased by a factor of two.

Contrary to this expectation, the higher-velocity model yields systematically lower magnetic field filling fractions and weaker large-scale magnetization at $z$\,$=$\,$0$ (albeit with very marginally boosted fractions at $z$\,$\sim$\,$3$). Inspection of the projected maps likewise reveals a reduction in the extent of strongly magnetized regions relative to the fiducial case. The origin of this behaviour is not the transport process itself, but rather the indirect impact of the modified wind prescription on galaxy evolution. In particular, altering the wind velocity leads to more pronounced ejection of gas away from the center of the halo, thereby reducing gas densities in the ISM. This ultimately changes the growth histories of galaxies and SMBHs, which in turn affects the timing and cumulative efficiency of subsequent magnetic energy injection.

This result highlights an important caveat when interpreting astrophysical magnetogenesis models. The resulting magnetic field distribution is not determined solely by the explicit transport properties of the outflows, but can also be sensitive to the manner in which those outflows couple to the broader galaxy formation model. Consequently, changes that appear favourable for magnetic transport in isolation do not necessarily lead to enhanced large-scale magnetization once the self-consistent evolution of galaxies and SMBHs is taken into account.

\subsection{Seeding smaller SMBHs at earlier times}\label{ssec:app_bh_seed}
In our fiducial setting, SMBHs are seeded once a FoF halo exceeds a threshold mass of $\sim$\,$7$\,$\times$\,$10^{10}$\,M${_\odot}$, with an initial seed mass of $\sim$\,$10^{6}$\,M${_\odot}$ (Section~\ref{sec:methods}). To explore the impact of the onset of SMBH activity, Figure~\ref{fig:app_bh_seed_mass} compares the fiducial model to a variation in which both the SMBH seed mass and the FoF halo mass threshold for seeding are reduced by a factor of two. This modification enables SMBHs to form in lower-mass halos and therefore begin injecting magnetic fields at earlier cosmic times.

For our specific parameter choices, the resulting changes in the magnetic field filling fractions are marginal but systematic. Earlier SMBH seeding leads to slightly elevated filling fractions at high redshift, consistent with the increased time available for magnetized outflows to enrich the surrounding medium. The projected magnetic field maps similarly show a more extended distribution of magnetized gas relative to the fiducial model.

Although the effect is smaller than those produced by variations to the transport or feedback prescriptions, the trend is encouraging in that it moves the predicted magnetization in the direction closer to the $z$\,$\sim$\,$3$ constraint. This suggests that the timing of SMBH formation constitutes an additional degree of freedom in astrophysical magnetogenesis models and may contribute, in combination with other model variations, to enhanced intergalactic magnetization at early times.

\end{document}